\documentclass[letter]{aa}

\usepackage{graphicx}
\usepackage{txfonts}
\usepackage{longtable}
\usepackage{subcaption}
\usepackage{lscape}
\usepackage{xcolor}
\usepackage{multirow}
\usepackage{array}
\usepackage{longtable}

\def\raa{RAA\ }

\def\Teff{\,T_{\mathrm{eff}}}
\def\logTeff{\,\log T_{\mathrm{eff}}}

\def\Lsun{\,L_{\odot}}
\def\Msun{\,M_{\odot}}
\def\Rsun{\,R_{\odot}}
\def\cd{\,\mathrm{c/d}}
\def\mmag{\,\mathrm{mmag}}
\def\logg{\,\log{g}}

\begin{document}

   \title{Discovery of resonating integration modes in triple-mode high-amplitude $\delta$ Scuti stars: A new evolutionary phase indicator}
\titlerunning{RI Mode in HADS}

   \author{Jia-Shu Niu
   \inst{1,2,3}\fnmsep\thanks{Corresponding Author, \email{jsniu@sxu.edu.cn}}
   \and Hui-Fang Xue
   \inst{4,5,6}\fnmsep
          }
   \institute{Institute of Theoretical Physics, Shanxi University, Taiyuan 030006, China;\\
        \and
        State Key Laboratory of Quantum Optics Technologies and Devices, Shanxi University, Taiyuan 030006, China;\\
        \and
             Collaborative Innovation Center of Extreme Optics, Shanxi University, Taiyuan 030006, China;\\
             \and
             Department of Physics, Taiyuan Normal University, Jinzhong 030619, China;\\
             \and
             Institute of Computational and Applied Physics, Taiyuan Normal University, Jinzhong 030619, China;\\
             \and
             Shanxi Key Laboratory for Intelligent Optimization Computing and Blockchain Technology, Jinzhong 030619, China;\\
             }

   \date{Received XX XX, 2025; accepted XX XX, 2025}

\abstract
{High-amplitude $\delta$ Scuti stars (HADS) that pulsate in their first three radial modes are rare in current samples. Here, we analyse five such triple-mode HADS observed by the Transiting Exoplanet Survey Satellite (TESS) and {report that the previously considered second overtone mode ($f_2$) is actually the non-radial component of a resonating integration mode (RI mode, resulting from resonance between a radial p-mode and a non-radial p-g mixed mode), which shows significant amplitude and frequency variations over short timescales (approximately 20 days).} This RI mode appears to be widespread among these stars. Notably, all five stars are in the post-main sequence evolutionary phase, actively crossing the Hertzsprung gap. These stars offer valuable insights into stellar evolution during the Hertzsprung gap, which is one of the most rapid evolutionary stages in a star's life.
}


   \maketitle

\section{Introduction}
\label{sec:intro}

$\delta$ Scuti stars are short-period pulsating variable stars with periods ranging from 15 minutes to 8 hours and spectral types A-F. They are located at the intersection of the main sequence (MS) and the lower section of the classical Cepheid instability strip on the Hertzsprung-Russell (H-R) diagram. The pulsations in $\delta$ Scuti stars are self-excited via the $\kappa$ mechanism, driven by partial ionisation of helium in their outer layers \citep{Kallinger2008,Handler2009,Guenther2009,Uytterhoeven2011,Holdsworth2014,Steindl2022}.

High-amplitude $\delta$ Scuti stars (HADS) are a sub-class characterised by larger amplitudes ($\Delta V \geq 0.1\ \mathrm{mag}$) and slower rotation speeds ($v \sin i \le 30\ \mathrm{km/s}$). However, as more HADS samples have been accumulated, these classical criteria have become less distinct \citep{Balona2012}. Most HADS pulsate with one or two radial modes \citep{Niu2013,Niu2017,Bowman2021,Daszynska2022}, while some exhibit three radial modes \citep{Wils2008,Niu2022,Xue2023}, four radial modes \citep{Pietrukowicz2013,Netzel2022a,Netzel2022b}, or even non-radial modes \citep{Poretti2011,Xue2020}.

To date, only 18 HADS in the Galaxy have been confirmed to pulsate in the first three radial modes -- the fundamental mode and the first two overtones ($f_0$, $f_1$, and $f_2$)\footnote{Pulsation mode identifications were based on period ratio relations \citep{Stellingwerf1979}} \citep{Xue2023}. With the increasing amount of high-precision photometric data from space telescopes (e.g. Kepler and TESS), unexpected pulsation properties have been discovered in these HADS.

\begin{figure*}[htbp!]
  \centering
  \includegraphics[width=0.9\textwidth]{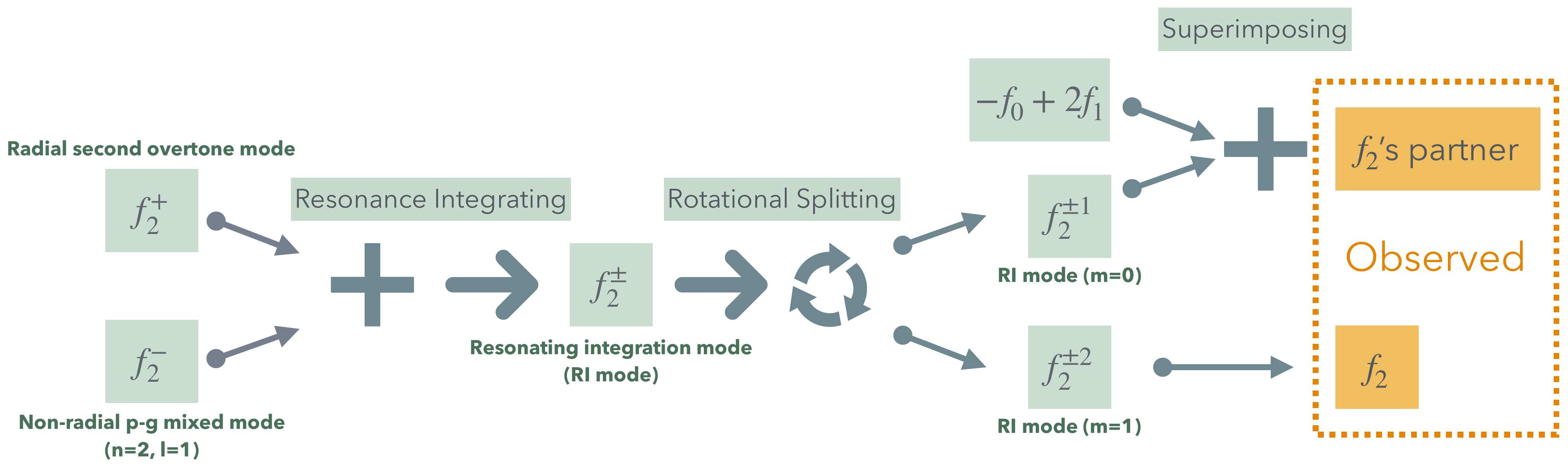}
  \caption{Framework based on the RI mode. The two elements in the orange dotted box represent frequencies directly observed in the frequency domain of light curves. The annotation of the labels are presented in bolded dark font.}
  \label{fig:RI_flowchart}
\end{figure*}

For example, in KIC 6382916, \citet{Ulusoy2013} and \citet{Mow2016} using Kepler long-cadence data found a significant increase in the amplitude of the second overtone mode ($f_2$). Subsequently, \citet{Niu2022} reported that the linear period variation rate of $f_2$ is about eight times larger than the ones of the first two modes ($f_0$ and $f_1$). Using Kepler short-cadence data, \citet{Niu2024} found that the harmonics of $f_0$ and $f_1$ exhibit amplitude and frequency variations uncorrelated with their parent modes, even at the first and second harmonics. Furthermore, significant interactions were observed between $f_2$ and its partner mode ($-f_0 + 2f_1$).

These unexpected pulsation behaviours can be naturally explained by the framework proposed in \citet{Niu2022}, based on a comprehensive study of KIC 6382916. In this framework, the third pulsation mode ($f_2$) is identified as the non-radial component of a resonating integration (RI) mode, generated by resonance between a radial p-mode and a non-radial p-g mixed mode.

Specifically, the RI mode framework is illustrated in Figure \ref{fig:RI_flowchart}: (i) resonance occurs between a radial p-mode ($f_{2}^{+}$, the radial second overtone) and a non-radial p-g mixed mode ($f_{2}^{-}$, with $n=2$ and $l=1$), producing a RI mode ($f_{2}^{\pm}$, RI mode); (ii) $f_{2}^{\pm}$, having both radial and non-radial properties, splits into two frequencies ($f_{2}^{\pm 1}$ with $m=0$ and $f_{2}^{\pm 2}$ with $m=1$) due to stellar rotation ($f_{2}^{\pm 2} - f_{2}^{\pm 1} = \Delta \omega$, where $\Delta \omega$ is the stellar rotational frequency); (iii) $f_{2}^{\pm 1}$ combines with the combination of $f_0$ and $f_1$ (i.e., $-f_0 + 2f_1$) to form the observed partner mode of $f_2$; and (iv) $f_{2}^{\pm 2}$ corresponds to the observed $f_2$. This framework not only explains the long-term amplitude increase and large frequency variation in $f_2$ but also predicts short-term features: (i) $f_2$ may exhibit amplitude or frequency modulation due to the similarity yet difference between $f_{2}^{+}$ and $f_{2}^{-}$; and (ii) the partner mode of $f_2$ ($-f_0 + 2f_1$) may show frequency splitting and asynchronous amplitude or frequency variations compared to $f_2$ due to its dual identity. Both short-term predictions were confirmed in \citet{Niu2024}, using Kepler short-cadence data.

It is now intriguing to investigate whether these phenomena are widespread in other radial triple-mode HADS. In this study, we examine the amplitude and frequency variations in $f_2$ and its partner mode ($-f_0 + 2f_1$) in five HADS stars: DO CMi, GSC 06047-00749 (ASAS J094303-1707.3), V0803 Aur, V1384 Tau, and V1393 Cen. These stars were previously identified as radial triple-mode HADS based on TESS space telescope time-series photometric data \citep{Xue2023}.

\section{Results}
\label{sec:results}

\begin{figure*}[htbp!]
  \centering
  \includegraphics[width=0.495\textwidth]{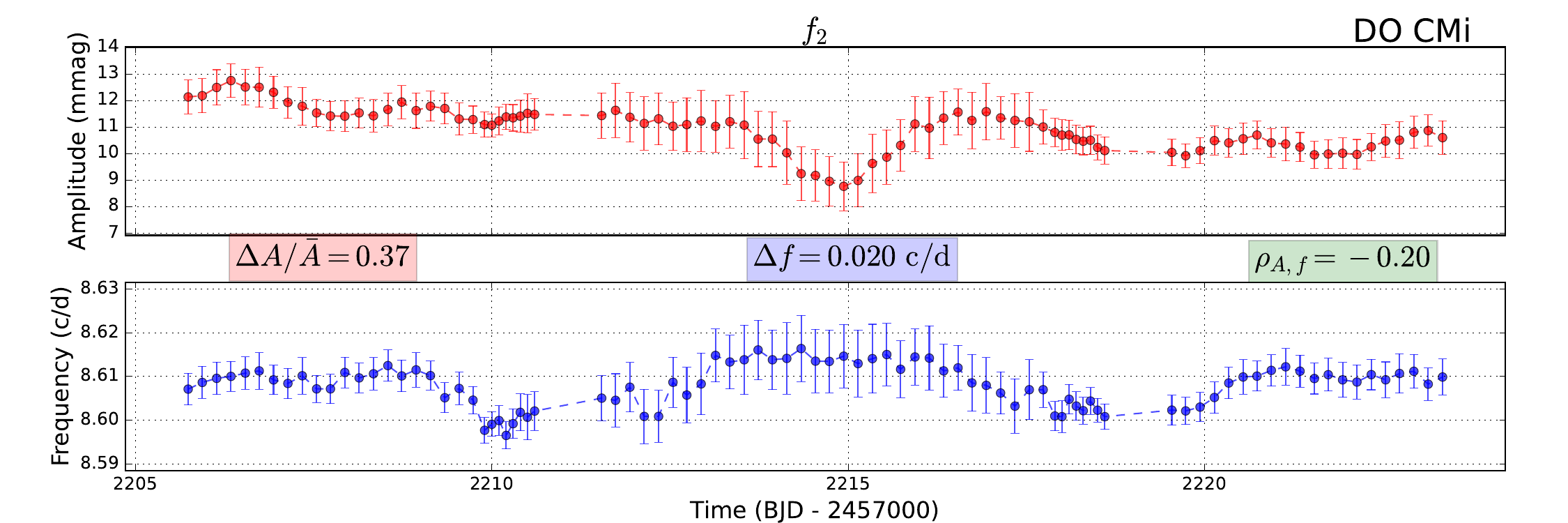}
  \includegraphics[width=0.495\textwidth]{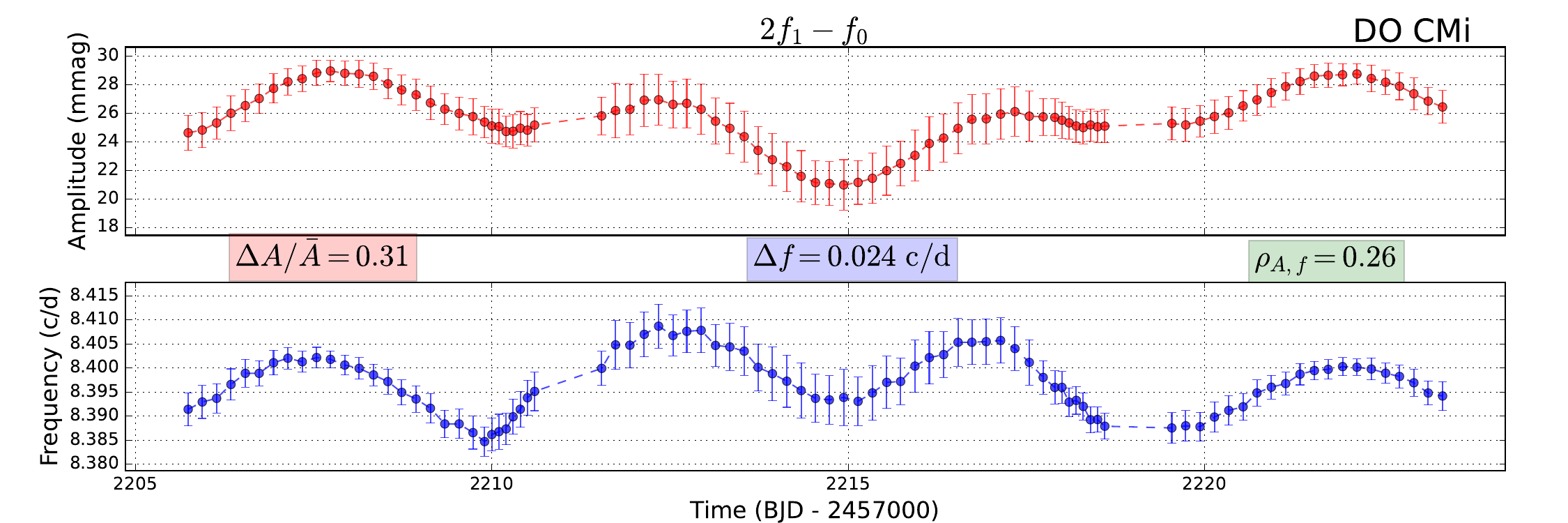}
  \includegraphics[width=0.495\textwidth]{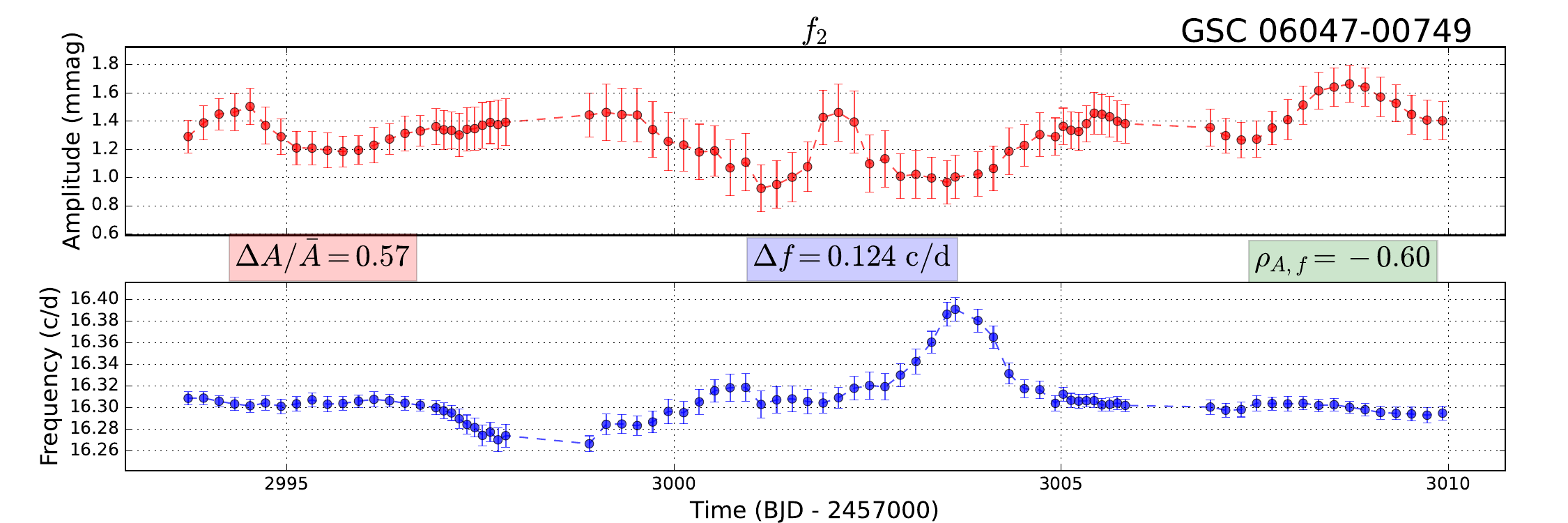}
  \includegraphics[width=0.495\textwidth]{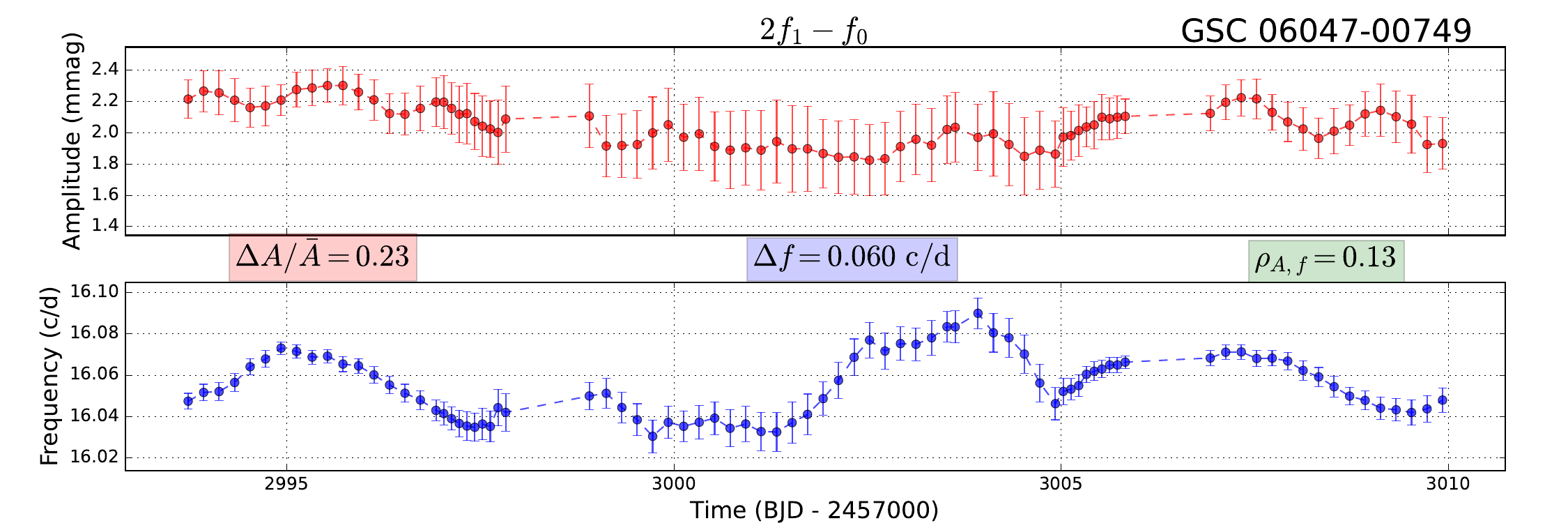}
  \includegraphics[width=0.495\textwidth]{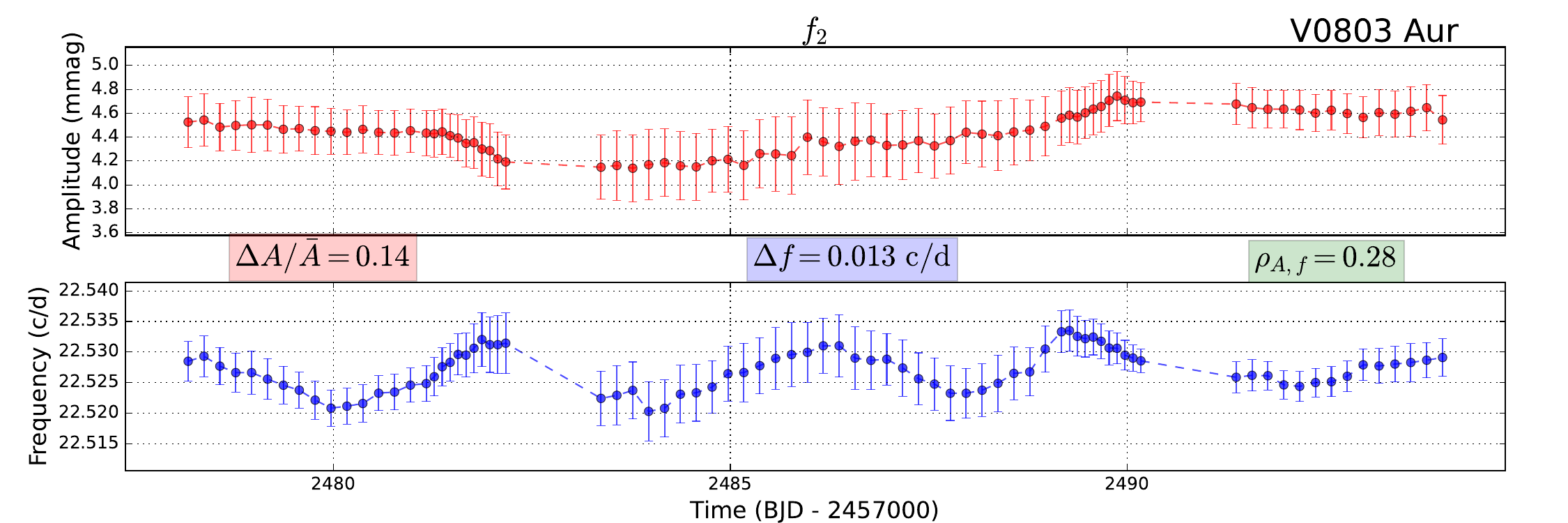}
  \includegraphics[width=0.495\textwidth]{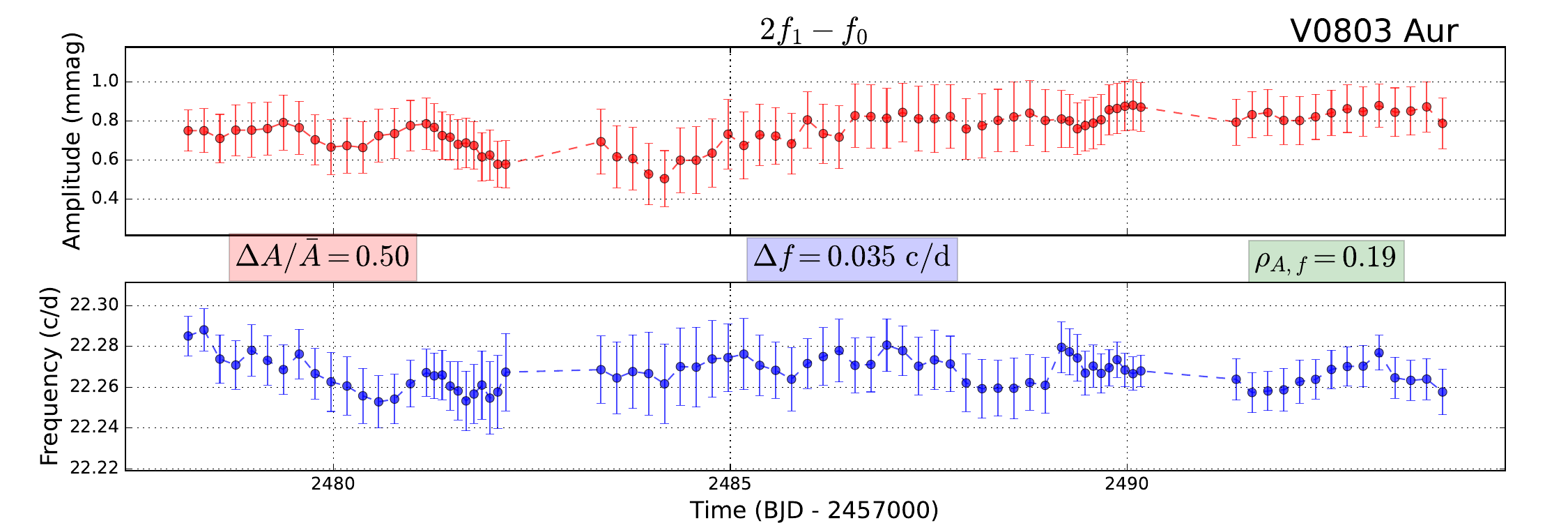}
  \includegraphics[width=0.495\textwidth]{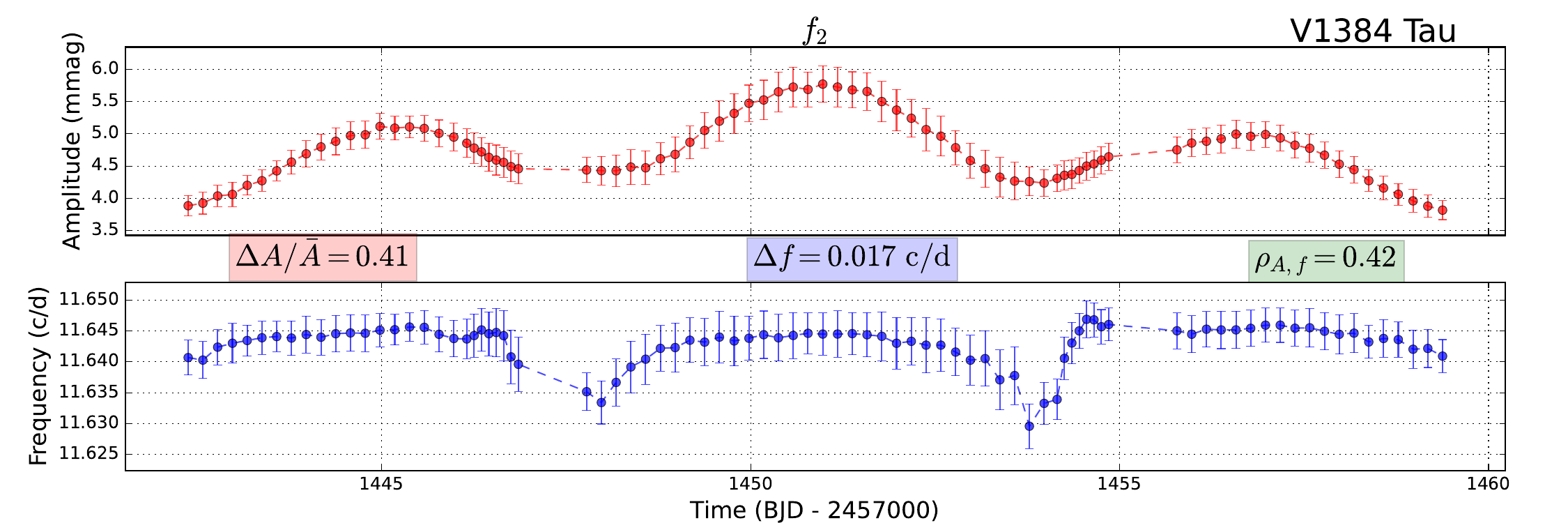}
  \includegraphics[width=0.495\textwidth]{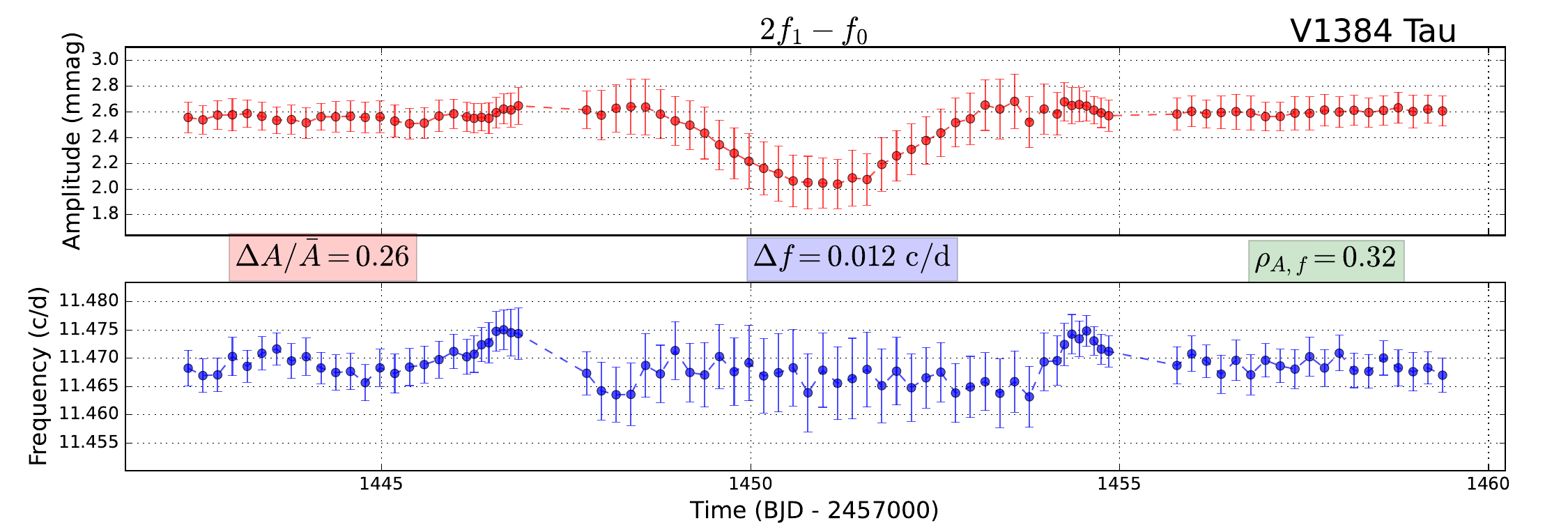}
  \includegraphics[width=0.495\textwidth]{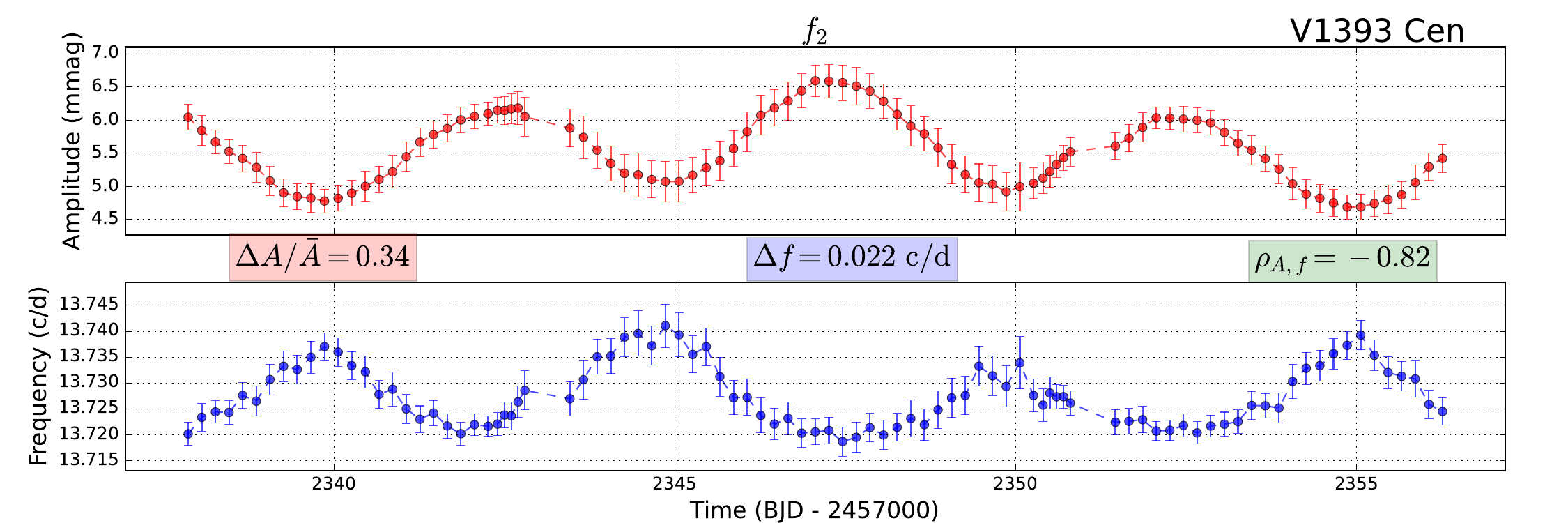}
  \includegraphics[width=0.495\textwidth]{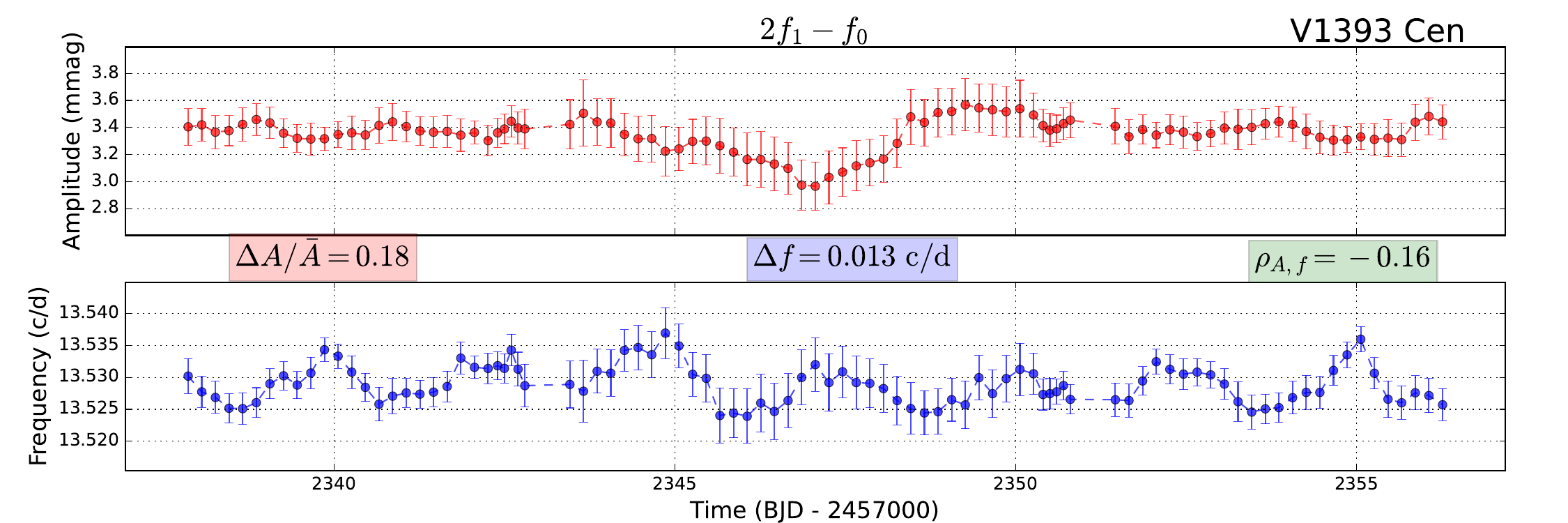}
  \caption{Amplitude and frequency variations in $f_2$ and its partner pulsation mode ($-f_0 + 2f_1$) for the five HADS. $\Delta A/\bar{A}$ is highlighted in red, which represents the relative amplitude variations. $\Delta f$ is highlighted in blue, which represents the absolute frequency variations. $\rho_{A,f}$ is highlighted in green, which represents a perfect correlation or anti-correlation between the variations in the amplitude and frequency when it equals +1 or -1 respectively.}
  \label{fig:5HADS}
\end{figure*}

Based on the same methods used in \citet{Xue2024}, we extracted the amplitude and frequency variations in $f_2$ and its partner pulsation mode ($-f_0 + 2f_1$) for the five HADS stars, as is shown in Figure \ref{fig:5HADS}. In Figure \ref{fig:5HADS}, the relative amplitude variations ($\Delta A/\bar{A} \equiv (A_\mathrm{max} - A_\mathrm{min})/\bar{A}$)\footnote{Here, $A_\mathrm{max}$ and $A_\mathrm{min}$ represent the maximum and minimum amplitude values within the moving windows, while $\bar{A}$ denotes the mean amplitude across these windows.}, the absolute frequency variations ($\Delta f \equiv f_\mathrm{max} - f_\mathrm{min}$)\footnote{$f_\mathrm{max}$ and $f_\mathrm{min}$ correspond to the maximum and minimum frequency values observed within the moving windows.}, and the Pearson correlation coefficient ($\rho_{A,f}$, which represents a perfect correlation or anti-correlation between the variations of the amplitude and frequency when it equals +1 or -1 respectively) between amplitude and frequency are displayed for a more quantitative representation.

As is illustrated in Figure \ref{fig:5HADS}, the amplitude and frequency of $f_2$ and $-f_0 + 2f_1$ in all five HADS exhibit significant short-term modulation, with amplitude variations ranging from $14\%$ to $57\%$ and frequency variations from $0.01$ to $0.06\ \mathrm{c/d}$ within 20 days. Furthermore, the amplitude and frequency variations in $f_2$ and $-f_0 + 2f_1$ are asynchronous in these HADS. Considering that they are locked by the rotation of the star (which should show synchronous frequency variations if $-f_0 + 2f_1$ were just the rotational splitting of $f_2$), this indicates that the partner pulsation mode of $f_2$ should have a dual identity. None of the amplitude and frequency variations in Figure \ref{fig:5HADS} show obvious (anti-)correlations, except for the $f_2$ of V1393 Cen, which exhibits a similar behaviour to the $f_2$ of KIC 6382916 \citep{Niu2024}.

\section{Discussions and conclusions}
\label{sec:dis_con}

It is quite interesting that the framework of the RI mode proposed in \citet{Niu2022} for KIC 6382916 can also be confirmed in other (previously considered) radial triple-mode HADS, indicating that the RI mode is quite prevalent in such stars. Due to the limited duration of the photometric data in this study, we can only confirm the short-term behaviour of $f_2$. With the accumulation of more photometric data for these HADS, we anticipate that $f_2$ will exhibit continuous amplitude increases or decreases and significant differences in linear period variation rates compared to $f_0$ and $f_1$ over long timescales, similar to the case of KIC 6382916.

Given that the RI mode arises from resonance between a radial p-mode and a non-radial p-g mixed mode (which corresponds to a specific structure), we predict that these stars should be in a specific evolutionary phase. Since $f_2$ is not the second overtone pulsation mode in the RI mode framework and always shows distinct frequency variations in the five HADS, we used $f_0$ and $f_1$ (which are confirmed as the fundamental and first overtone pulsation modes with relatively stable frequencies) to determine the evolutionary phase of these HADS.

As was done in \citet{Niu2022}, the asteroseismological models of the five HADS were constructed using the Modules for Experiments in Stellar Astrophysics (MESA; \citep{Paxton2011, Paxton2013, Paxton2015, Paxton2018, Paxton2019, Jermyn2023}) and the stellar oscillation code GYRE \citep{Townsend2013,Townsend2018,Goldstein2020,Sun2023}. All metallicity values (listed in Table \ref{tab:best_models}) are from Gaia DR3 \citep{Gaia2022}. Further details of the theoretical calculations can be found in Appendix \ref{app:02}.

The evolutionary tracks for each of the five HADS were calculated from the pre-MS to the red giant (RG) branch with a mass step of $0.01 \Msun$, covering a mass interval of $1 \Msun$ (depending on the best-fit models). Based on the pulsation frequencies calculated at each step along the evolutionary tracks, we obtained the best-fit seismic models (with the smallest $\chi^2$ values relative to the observed $f_0$ and $f_1$) for the five HADS (see Table \ref{tab:best_models} for details). The H-R and Petersen diagrams are shown in Figures \ref{fig:HR} and \ref{fig:petersen}, respectively.

\begin{table*}[hbtp!]
\centering
  \caption{Information on the best-fit seismic models.}
  \label{tab:best_models}
 \resizebox{1.0\textwidth}{!}{
  \begin{tabular}{l|cc|cc|cc|cc|cc}
    \hline
    \hline
  ID & \multicolumn{2}{c}{DO CMi} \vline& \multicolumn{2}{c}{GSC 06047-00749} \vline & \multicolumn{2}{c}{V0803 Aur} \vline& \multicolumn{2}{c}{V1384 Tau} \vline& \multicolumn{2}{|c}{V1393 Cen} \\ \hline
  & mod & obs & mod & obs & mod & obs & mod & obs & mod & obs \\
    \hline
  $f_0$ ($\cd$)   & 5.1421 & 5.141(6) & 10.0806 & 10.082(9) & 14.0714 & 14.073(5) & 7.1556 & 7.153(4) & 8.4906 & 8.490(2) \\ 
  $f_1$ ($\cd$)   &  6.7282 & 6.728(4) & 13.0725 & 13.069(0) & 18.1743 & 18.171(6) & 9.3108 & 9.311(7) & 11.0092 & 11.009(3) \\ 
  \hline
  $f_2$ ($\cd$)   & --- & 8.607(8) & --- & 16.30(3) & --- & 22.527(2) & --- & 11.643(0) & --- & 13.727(1)\\
  $2f_1 - f_0$ ($\cd$) & --- & $8.397(3)$ & --- & $16.054(6)$ & --- & $22.27(0)$ & --- & $11.469(5)$  & --- & $13.528(3)$  \\
  $\Delta \omega$ ($\cd$) & --- & $0.210(5)$ & --- & $0.24(8)$ & --- & $0.25(7)$ & --- & $0.173(5)$  & --- & $0.198(8)$  \\
  $f_2^{+}$ ($\cd$)   & {8.4127} & {---} & {16.3373} & {---} & {22.7402} & {---} & {11.6408} & {---} & {13.7258} & {---}\\
  \multirow{2}{*}{$f_2^{-}$ ($\cd$)}   & {8.3548} & {---} & {15.5752} & {---} & {21.4739} & {---} & {11.4123} &{---} &{13.5349} & {---}\\
   & (1,2,31)$^{*}$ & --- & (1,2,15) & --- & (1,2,10) & --- & (1,2,30) & --- & (1,2,23) & --- \\
  \hline
  $\log \Teff$ & 3.805 & [3.834,3.883] & 3.814 & [3.880,3.883] & 3.816 & [3.853,3.859] & 3.807 & [3.820,3.845] & 3.812 & [3.855,3.857] \\
  $\log L/\Lsun$ & 1.343 & --- & 0.904 & --- & 0.639 & --- & 1.114 & --- & 1.018 & --- \\
  $\logg$ & 3.492 & [3.476,3.545] & 3.833 & [3.889,3.921] & 3.985 & [3.934,4.010] & 3.657 & [3.647,3.706] & 3.739 & [3.719,3.729] \\
  $\log R/\Rsun$ & 0.584 & --- & 0.348 & --- & 0.211 & --- & 0.465 & --- & 0.407 & --- \\
  Age (Gyr) & 1.55 & --- & 3.33 & --- & 7.43 & --- & 2.24 & --- & 2.77 & --- \\
  Mass ($\Msun$) & 1.67 & --- & 1.23 & --- & 0.93 & --- & 1.41 & --- & 1.30 & --- \\
  \hline
  Z & 0.008 & [0.0079,0.0079] & 0.0046 & [0.0043,0.0050] & 0.0010 & [0.0006,0.0016] & 0.0054 & [0.0047,0.0068] & 0.004 & [0.0035,0.0038] \\
  $\mathrm{[Fe/H]}$ & -0.2223 & [-0.2253,-0.2210] & -0.4629  & [-0.4972,-0.4277] & -1.1400 & [-1.3924,-0.9232] & -0.3970 & [-0.4583,-0.2890] & -0.5545 & [-0.5874,-0.5468] \\
\hline
\end{tabular}
}
\footnotesize{Note: 'mod' represents the values obtained by (or used in) the theoretical best-fit models, and 'obs' represents the values from observations; all the observed values of $\logTeff$, $\logg$, and $\mathrm{[Fe/H]}$ come from Gaia DR3 \citep{Gaia2022}; $Z$ values were obtained based on the $\mathrm{[Fe/H]}$ values; $f_2^{+}$ denotes the frequency of the second overtone pulsation mode, and $f_2^{-}$ denotes the frequency of the p-g mixed mode, which is assumed to resonate with $f_2^{+}$; $^{*}$: the numbers in parentheses  ($l$, $n_p$, $n_g$) denote $l$, $n_p$, and $n_g$, respectively. All the pulsation modes from theoretical best-fit models listed here are unstable modes according to the GYRE non-adiabatic calculation.}
\end{table*}

Furthermore, it is noteworthy that all five HADS are located in the {post-MS} evolutionary phase, with helium cores and hydrogen-burning shells, actively crossing the Hertzsprung Gap. This finding aligns with our prediction and suggests that the (previously considered) radial triple-mode HADS should be in a special evolutionary phase, in which $f_2$ is no longer the radial second overtone pulsation mode but the non-radial component of the RI mode.

Generally, HADS are considered to be in MS and post-MS evolutionary phases. However, confirming the evolutionary phase of a single HADS via asteroseismology is challenging. In recent years, the evolutionary phases of some radial double-mode HADS have been confirmed using asteroseismology, including the MS phase \citep{Bowman2021,Chen2025} and post-MS phase (which can be divided into the overall contraction phase \citep{Daszynska2022,Daszynska2023} and hydrogen-burning shell phase \citep{Niu2017,Xue2018,Xue2022,Daszynska2023,Daszynska2024} in detail).

Figure \ref{fig:profile} presents the radial displacement (represented by $r \rho^{1/2} \delta r$) of the related pulsation modes (the second overtone and the p-g mixed mode) and some relevant profiles (including the sound speed (csound), opacity, convection velocity (convection), and hydrogen and helium fractions (H fraction and He fraction)) of the best-fit seismic models for the five HADS. The radial displacement of the second overtone and the p-g mixed modes show close nodes and similar phases along with the radius in the outer layers, which provides a necessary condition for resonance. The hydrogen and helium profiles and the convection region indicate that the star has a helium core, left the MS, and entered the post-MS evolutionary phase. All the sound speed profiles of the five stars show a similar structure near the core region, jumping near the edges of the cores and then beginning to decrease steadily.
Considering that the the convection region will extend into the deeper region of the stars and the $\kappa$ mechanism will be suppressed in the RG phase, we could conclude that these type of HADS are located in a transition phase with a hydrogen burning shell before the RG phase. Based on this work and \citet{Niu2022}, we can expect that the (previously considered) radial triple-mode HADS are in the post-MS phase with helium cores and hydrogen-burning shells.

In summary, all five (previously considered) radial triple-mode HADS studied in this work exhibit a distinct amplitude and frequency modulation in $f_2$, which was previously considered the second overtone pulsation mode but is now identified as the non-radial component of an RI mode. It appears that the RI mode is a common phenomenon in these triple-mode HADS. Moreover, all these triple-mode HADS are located in the post-MS evolutionary phase, crossing the Hertzsprung Gap, which reveals interesting relationships between their pulsation behaviours and evolutionary properties. Although whether all (previously considered) radial triple-mode HADS are in the same evolutionary phase (post-MS) requires further confirmation, the confirmed cases (KIC 6382916 \citep{Niu2022} and the five HADS in this work) provide excellent samples for studying stellar evolution in the Hertzsprung Gap, one of the most rapidly evolving stages in a star's life.

\begin{acknowledgements}
We would like to thank Jue-Ran Niu for providing us with an efficient working environment. H.F.X. acknowledges support from the National Natural Science Foundation of China (NSFC) (No. 12303036).
The authors acknowledge the TESS Science team and everyone who has contributed to making the TESS mission possible. 
\end{acknowledgements}

\clearpage
\setcounter{figure}{0}
\setcounter{table}{0}
\renewcommand{\thefigure}{A\arabic{figure}}
\renewcommand{\thetable}{A\arabic{table}}
\onecolumn
\begin{appendix}

\section{Overview of the important frequencies and data of the five stars}
\label{app:01}

Fourier analysis was applied to the light curves, and the amplitudes and frequencies of $f_0$, $f_1$, and $f_2$ are presented in Table \ref{tab:basic_info}.

\begin{table*}[hbtp!]
\centering
  \caption{Amplitudes and frequencies of the $f_0$, $f_1$, and $f_2$ pulsation mode of the five HADS.}
  \label{tab:basic_info}
  \resizebox{1.0\textwidth}{!}{
  \begin{tabular}{l|c|c|c|c|c}
    \hline
    \hline
  ID & DO CMi & GSC 06047-00749 & V0803 Aur & V1384 Tau & V1393 Cen \\
    \hline
  $A_0$ ($\mmag$) & $49\pm 1$ & $53\pm 1$ & $49.5\pm 0.6$ & $52.8\pm 0.8$ & $41.3\pm 0.5$ \\ 
  $f_0$ ($\cd$)   & $5.1416\pm 0.0005$ & $10.0829\pm 0.0004$ & $14.0735\pm 0.0003$ & $7.1534\pm 0.0003$ & $8.4902\pm 0.0003$ \\ 
  $A_1$ ($\mmag$) & $49.0\pm 0.7$ & $61\pm 1$ & $20.8\pm 0.2$ & $49.9\pm 0.5$ & $71\pm 1$ \\ 
  $f_1$ ($\cd$)   & $6.7284\pm 0.0003$ & $13.0690\pm 0.0005$ & $18.1716\pm 0.0003$ & $9.3117\pm 0.0002$ & $11.0093\pm 0.0003$ \\ 
  $A_2$ ($\mmag$) & $11.6\pm 0.2$ & $1.42\pm 0.07$ & $4.5\pm 0.1$ & $4.41\pm 0.07$ & $5.65\pm $  0.09 \\ 
  $f_2$ ($\cd$)   & $8.6078\pm 0.0004$ & $16.303\pm 0.001$ & $22.5272\pm 0.0005$ & $11.6430\pm 0.0003$ & $13.7271\pm 0.0003$ \\
  \hline
\multirow{4}{*}{TESS Sector} & {07$^{d}$ (24 d), 33$^{*,c}$ (26 d)} & {08$^{a}$ (24 d), 35$^{a}$ (24 d),} & {43$^{*,a}$ (24 d), 44$^{a}$ (24 d),} & {05$^{*,a}$ (25 d), 70$^{b}$ (25 d),} & {11$^{a}$ (25 d), 38$^{*,a}$ (27d), } \\
 & {87$^{b}$ (26 d)} &  {62$^{*,b}$ (24 d)} &{45$^{a}$ (25 d), 60$^{b}$ (26 d),} & {71$^{b}$ (26 d)} & {65$^{b}$ (27 d)}\\
  & &  &{71$^{b}$ (26 d), 72$^{b}$ (25 d),} & & \\
  & &  &{73$^{b}$ (27 d)} & & \\
  \hline
\end{tabular}
}
\\
\footnotesize{Note: $^{*}$ denotes the data used in this work; $^{a}$, $^{b}$, $^{c}$, and $^{d}$ denote the data cadence of 120 s, 200 s, 600 s, and 1800 s, respectively; the number in parentheses denotes the length of the data set in days.}
\end{table*}

\section{Theoretical model calculation}
\label{app:02}

In order to determine the stellar mass and evolutionary phase based on the single star evolutionary models, the open source 1D stellar evolution code Modules for Experiments in Stellar Astrophysics (MESA, r24.08.1, \citet{Paxton2011,Paxton2013,Paxton2015,Paxton2018,Paxton2019,Jermyn2023}) was used to construct the structural and evolutionary models. At each step along with the evolutionary tracks, the pulsation frequencies of the specific structure were calculated by the stellar oscillation code GYRE \citep{Townsend2013,Townsend2018,Goldstein2020,Sun2023}.
All metallicity values (listed in Table \ref{tab:best_models}) are from Gaia DR3 \citep{Gaia2022}.

The following formulas were used to calculate the initial heavy element abundance $Z$, initial hydrogen abundance $X$, and the helium abundance $Y$:
\begin{equation}
  \mathrm{[Fe/H]} = \log \frac{Z}{X} - \log \frac{Z_\odot}{X_\odot} ,
\end{equation}
\begin{equation}
  \label{equ:Y(Z)}
  Y=0.24 + 3Z ,
\end{equation}
\begin{equation}
  X+Y+Z=1 ,
\end{equation}
where $X_\odot =0.7381$ and $Z_\odot =0.0134$ \citep{Asplund2009}. Eq.(\ref{equ:Y(Z)}) was provided by \citet{Mowlavi1998}. 

At the same time, the values of the option {\tt Zbase} (which provides the reference metallicity necessary to calculate element variations) were set to be the initial metallicity of the stars, and {\tt initial\_zfracs = 'AGSS09\_zfracs'} was selected. For the opacity, {\tt kap\_file\_prefix = 'a09'}, {\tt kap\_CO\_prefix = 'a09\_co'} and {\tt kap\_lowT\_prefix = 'lowT\_fa05\_a09p'} were selected.

The the mixing-length parameter $\alpha_{\mathrm{MLT}} = 1.89$ \citep{Niu2017,Niu2022,Xue2023} was used all through this work; the exponential scheme overshooting \citep{Herwig2000} whose parameter depends on the stellar mass $M$ (in solar masses): $f_{\mathrm{ov}} = (0.13 M - 0.098)/9.0$ \citep{Magic2010}) was adopted from pre-MS to post-MS with a mass step of 0.01 $\Msun$ covering a mass interval of 1 $\Msun$ (depending on the best-fit models).

The rotation of the star was not considered in the model calculation, because of the relative slow rotation of the star ($\Delta \omega < 0.3\ \cd$). 
All the evolutionary tracks were calculated from the pre-MS to the RG branch.

Based on the pulsation frequencies calculated in every step along with the evolutionary tracks, we got the best-fit seismic models (which have the smallest $\chi^2$ with respect to the observed values of $f_0$ and $f_1$) for all the five stars (see in Table \ref{tab:best_models} for detailed information). The H-R diagram and Petersen diagram are shown in Figure \ref{fig:HR} and Figure \ref{fig:petersen}, respectively.

\section{Figures}
\label{app:03}

\begin{figure*}[htbp!]
  \centering
  \includegraphics[width=0.49\textwidth]{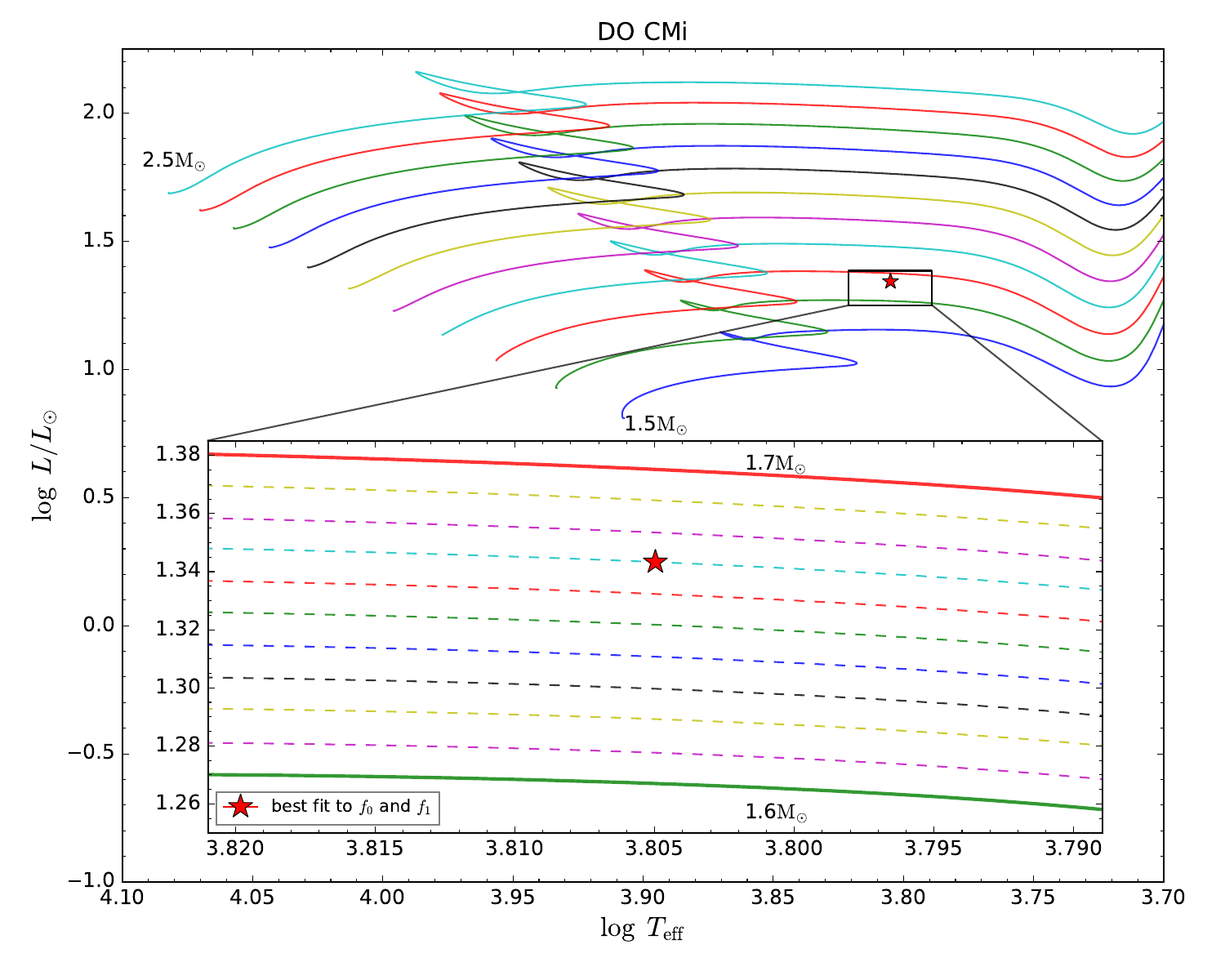}
  \includegraphics[width=0.49\textwidth]{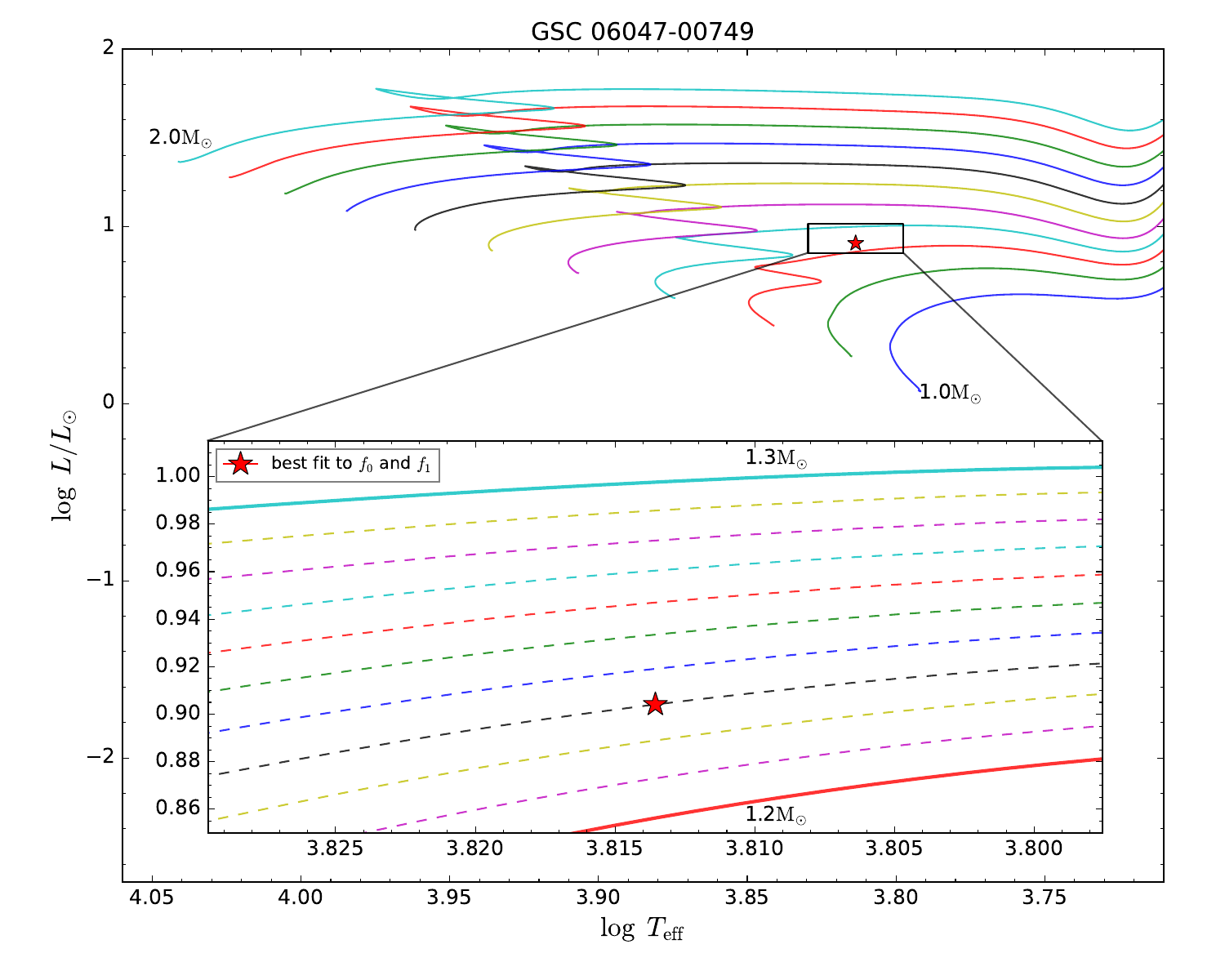}
  \includegraphics[width=0.49\textwidth]{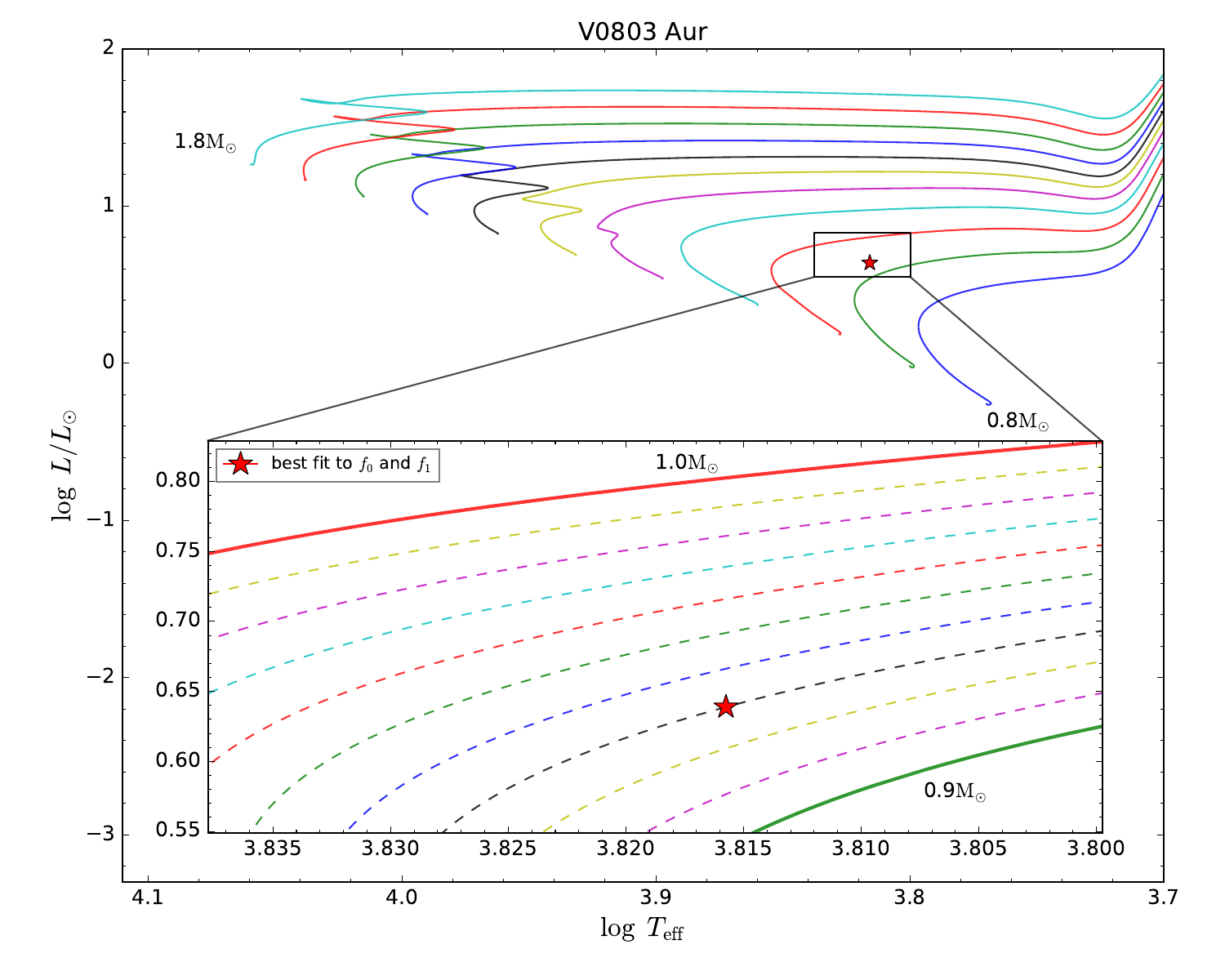}
  \includegraphics[width=0.49\textwidth]{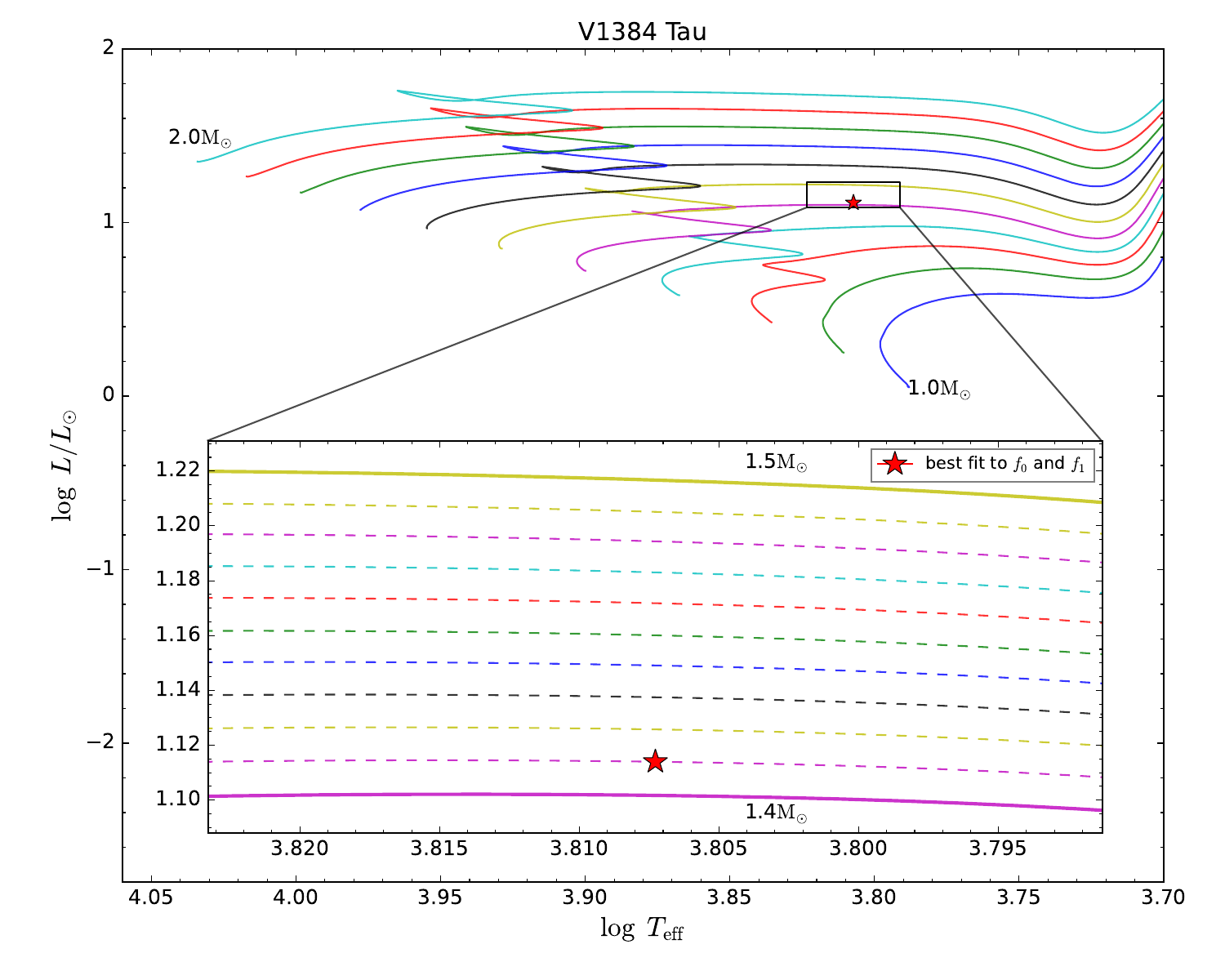}
  \includegraphics[width=0.49\textwidth]{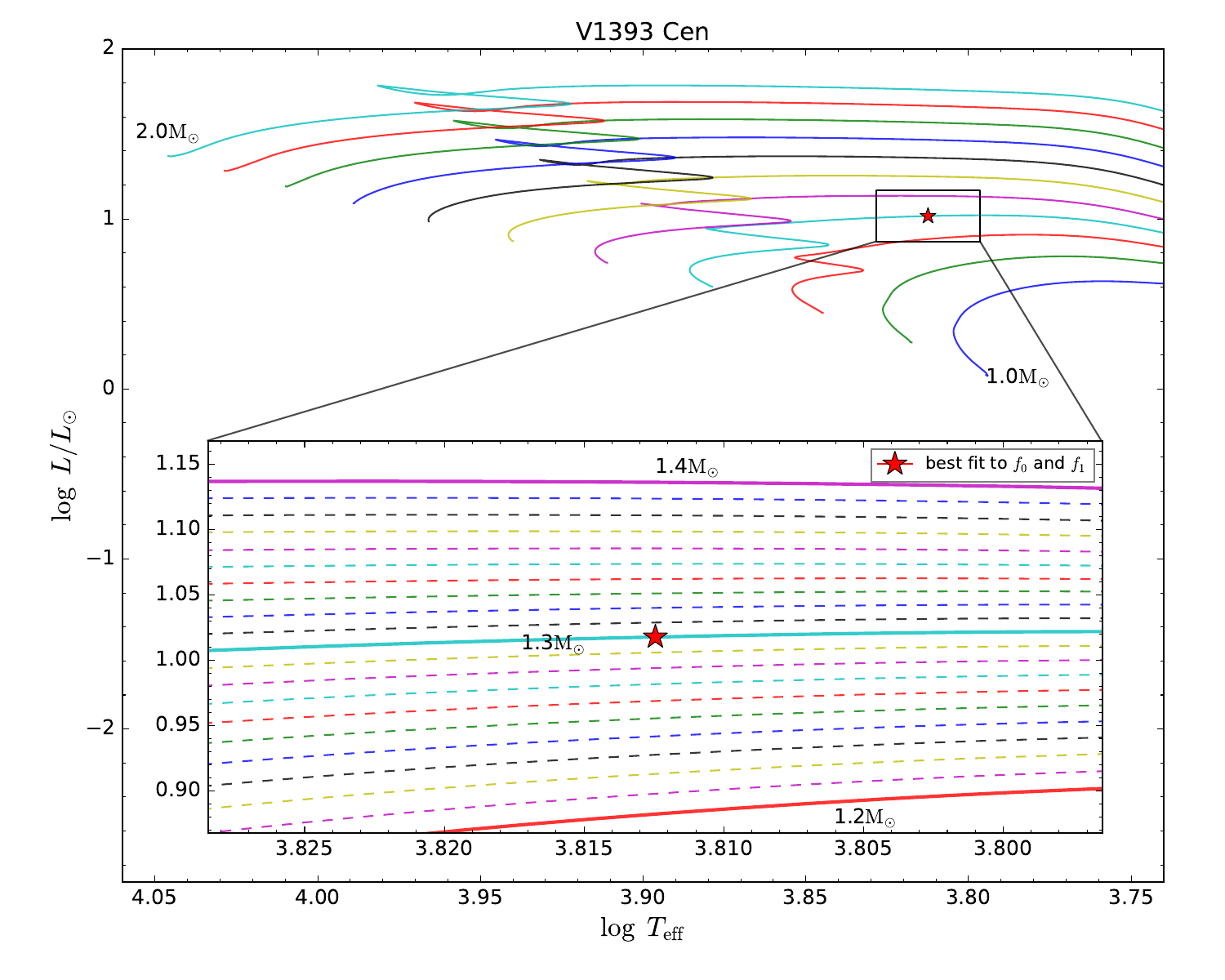}
  \caption{Hertzsprung-Russell diagram with all the evolutionary tracks and the best-fit seismic models of the five HADS. The coloured solid evolutionary tracks show the stages from the zero-age main sequence (ZAMS) to the post-main sequence (post-MS) evolutionary phase, with initial masses covering $1 \Msun$. The regions surrounded by the black rectangular boxes are zoomed-in views of the best-fit seismic models, represented by red stars. The coloured dashed lines indicate evolutionary tracks with initial mass steps of $0.01 \Msun$.}
  \label{fig:HR}
\end{figure*}

\begin{figure*}[htp]
  \centering
  \includegraphics[width=0.49\textwidth]{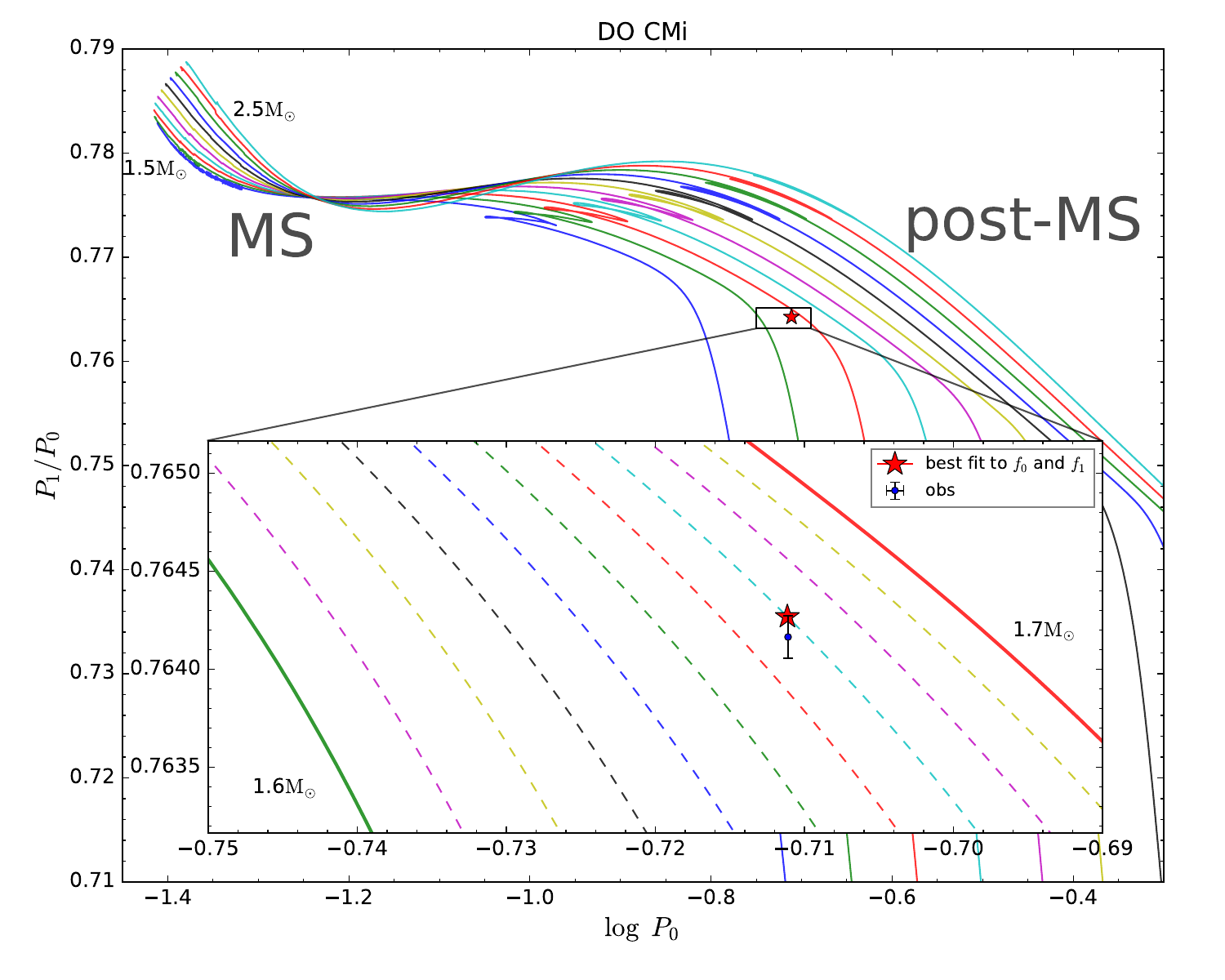}
  \includegraphics[width=0.49\textwidth]{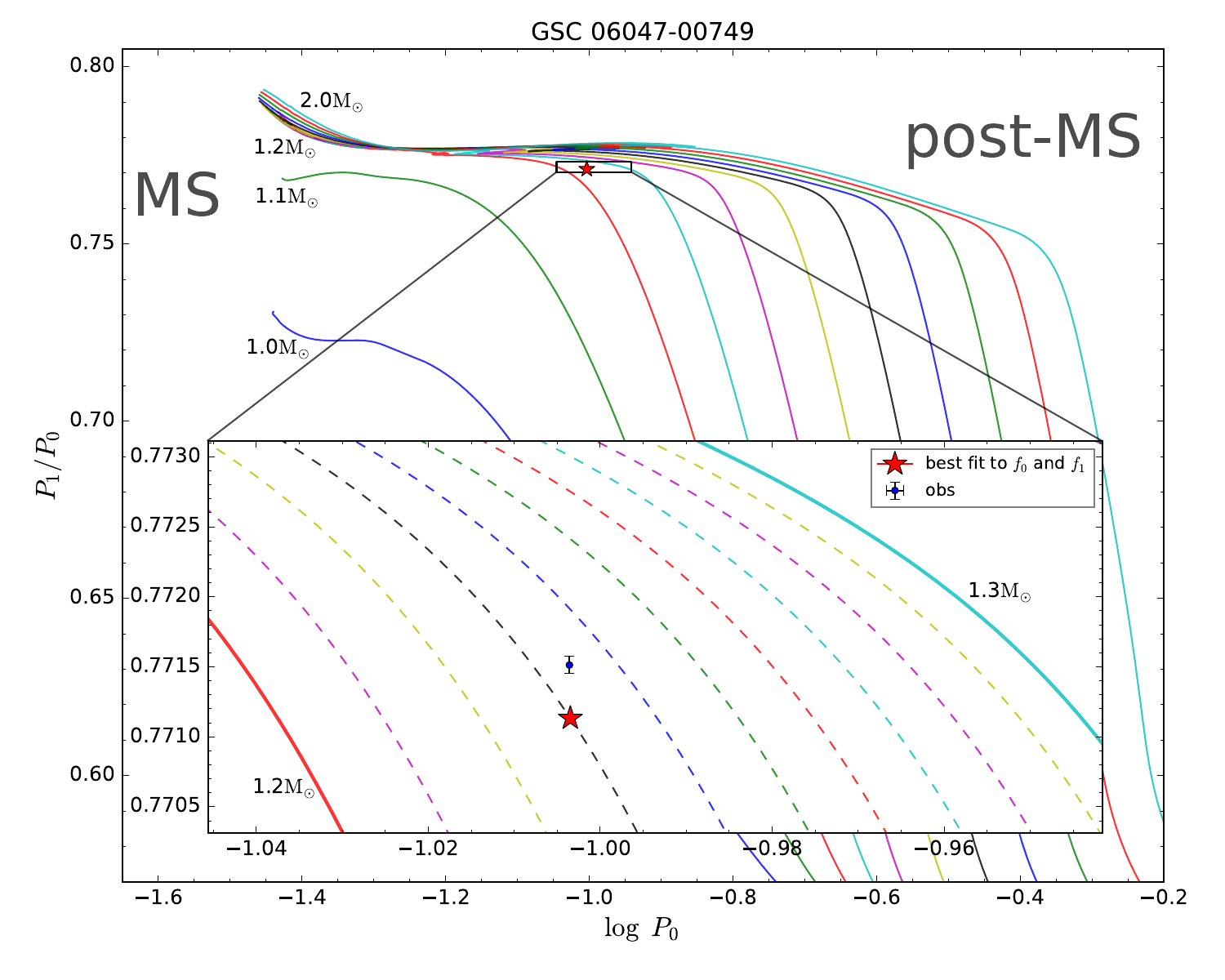}
  \includegraphics[width=0.49\textwidth]{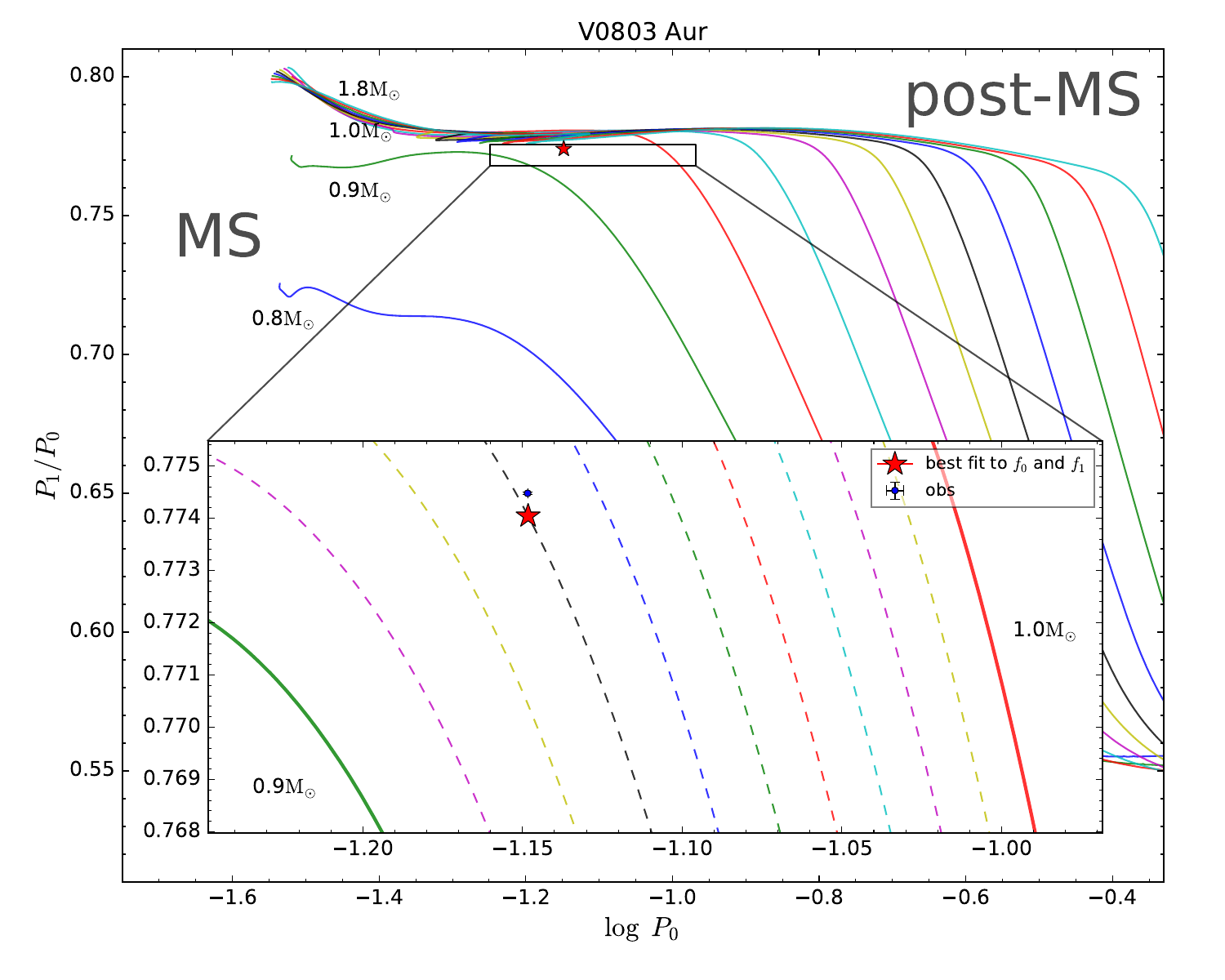}
  \includegraphics[width=0.49\textwidth]{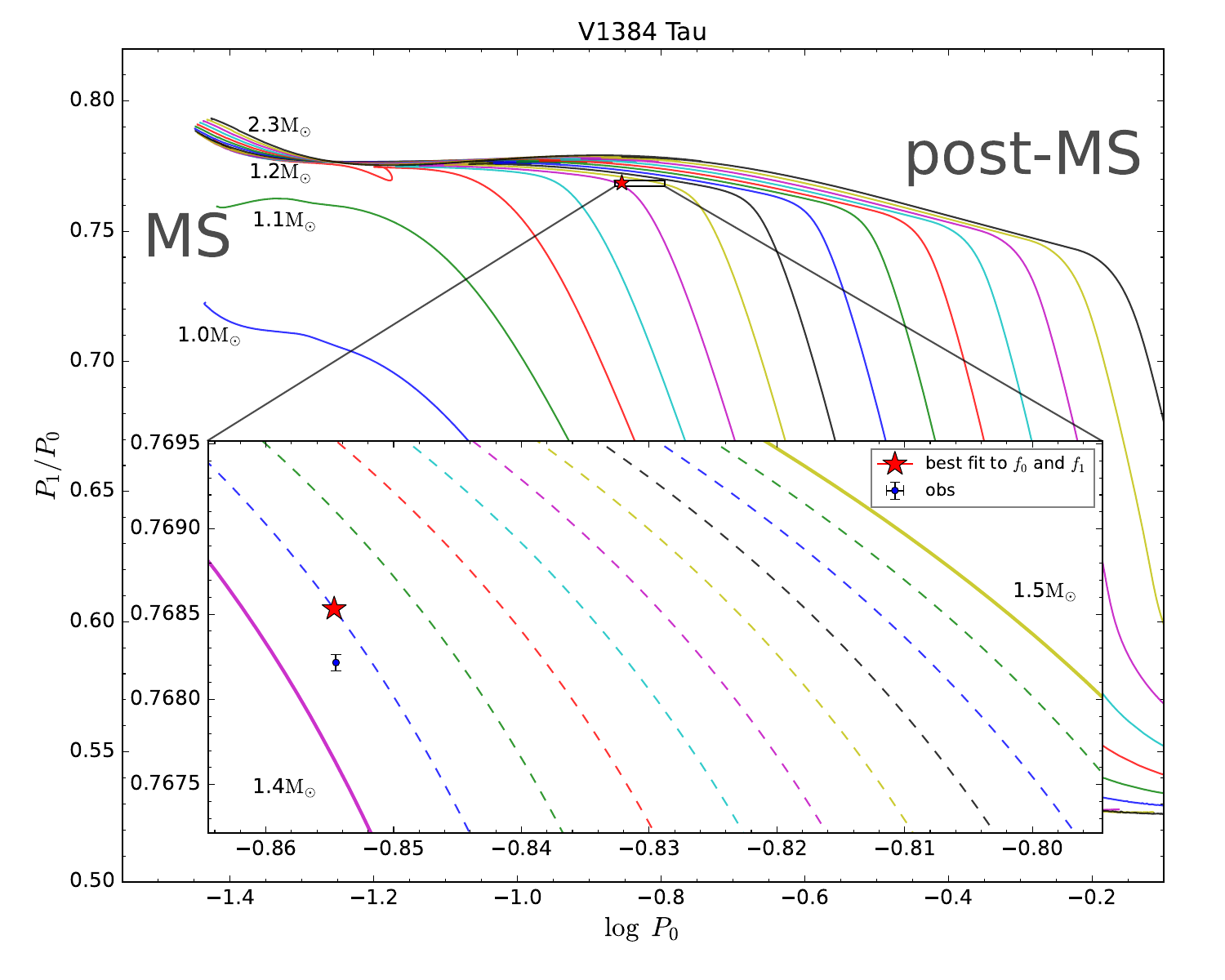}
  \includegraphics[width=0.49\textwidth]{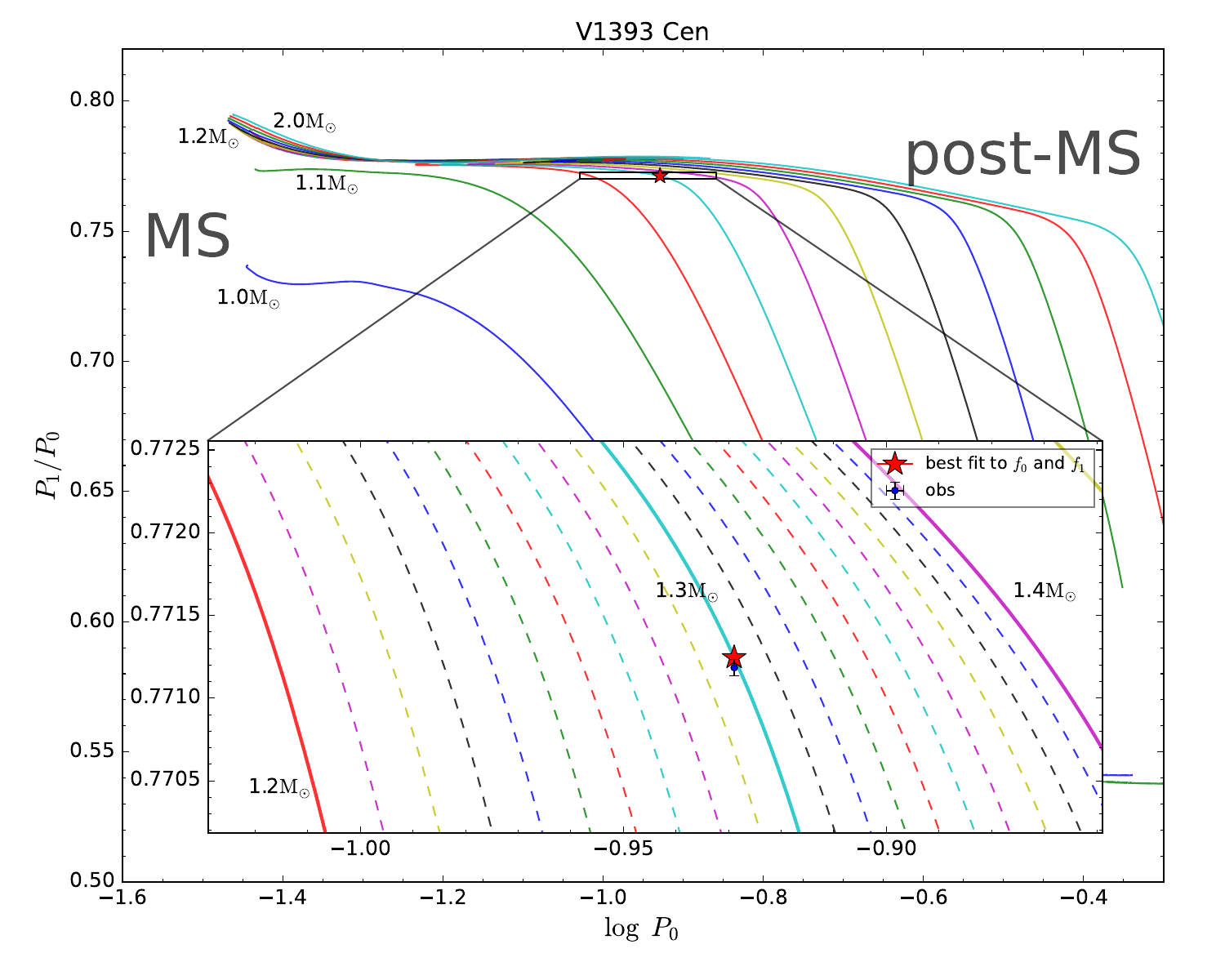}
  \caption{Petersen diagram with all evolutionary tracks and best-fit seismic models of the five HADS. The coloured solid evolutionary tracks show the stages from the zero-age MS to the post-MS evolutionary phase, with initial masses covering $1 \Msun$. The regions surrounded by the black rectangular boxes are selected to zoom in the best-fit seismic models, which are represented by the red stars. The coloured dashed lines present the evolutionary tracks with initial mass steps of $0.01 \Msun$.}
  \label{fig:petersen}
\end{figure*}

\begin{figure*}[htp]
  \centering
  \includegraphics[width=0.49\textwidth]{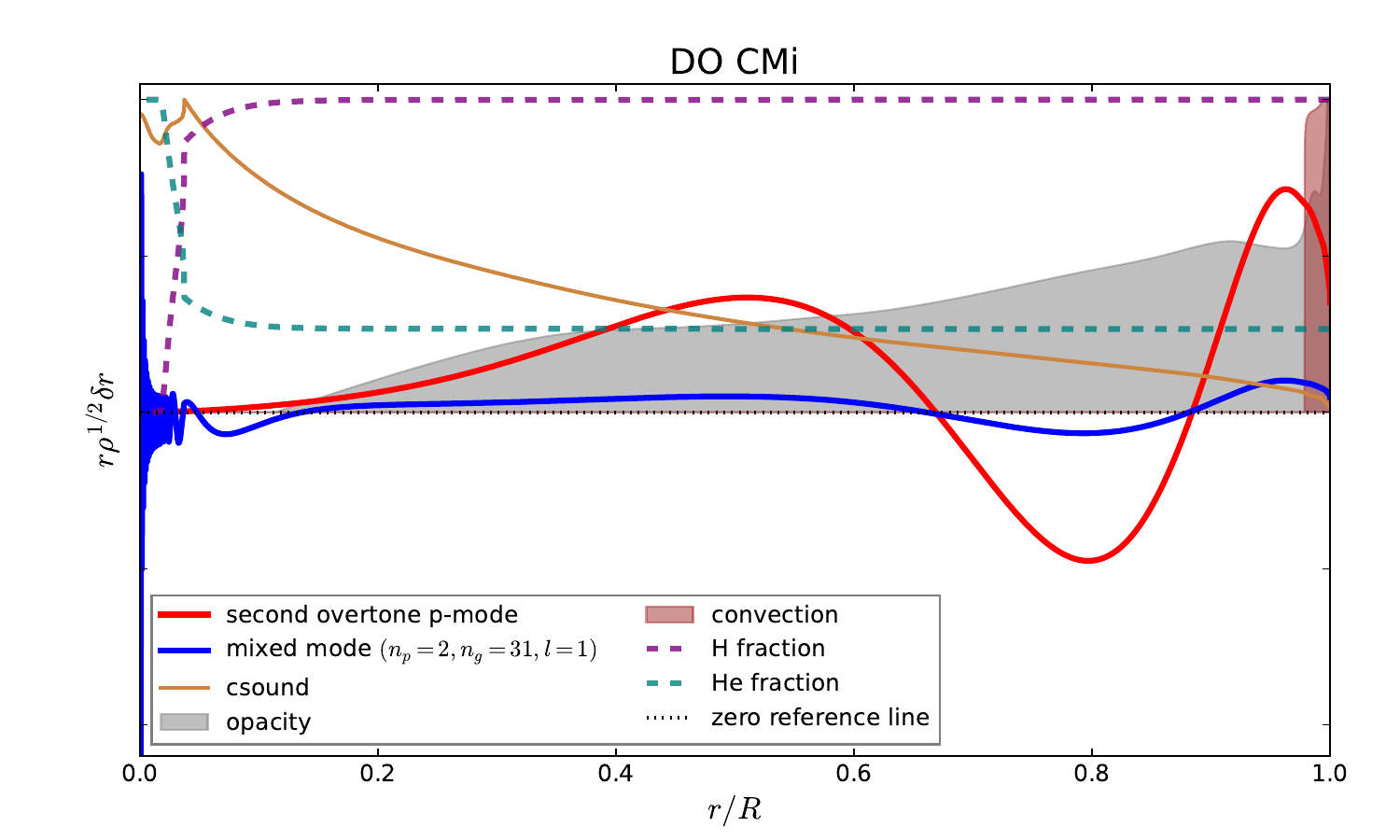}
  \includegraphics[width=0.49\textwidth]{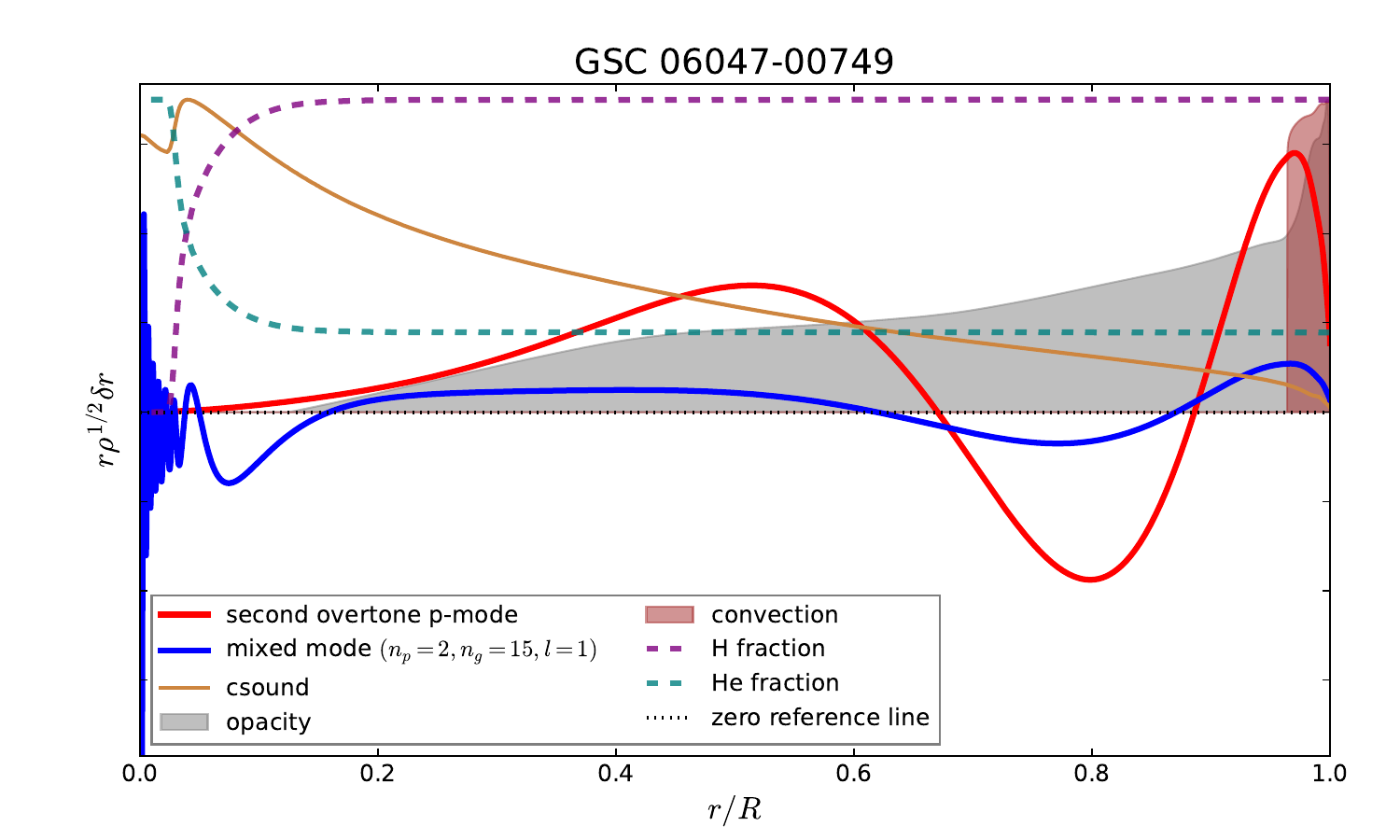}
  \includegraphics[width=0.49\textwidth]{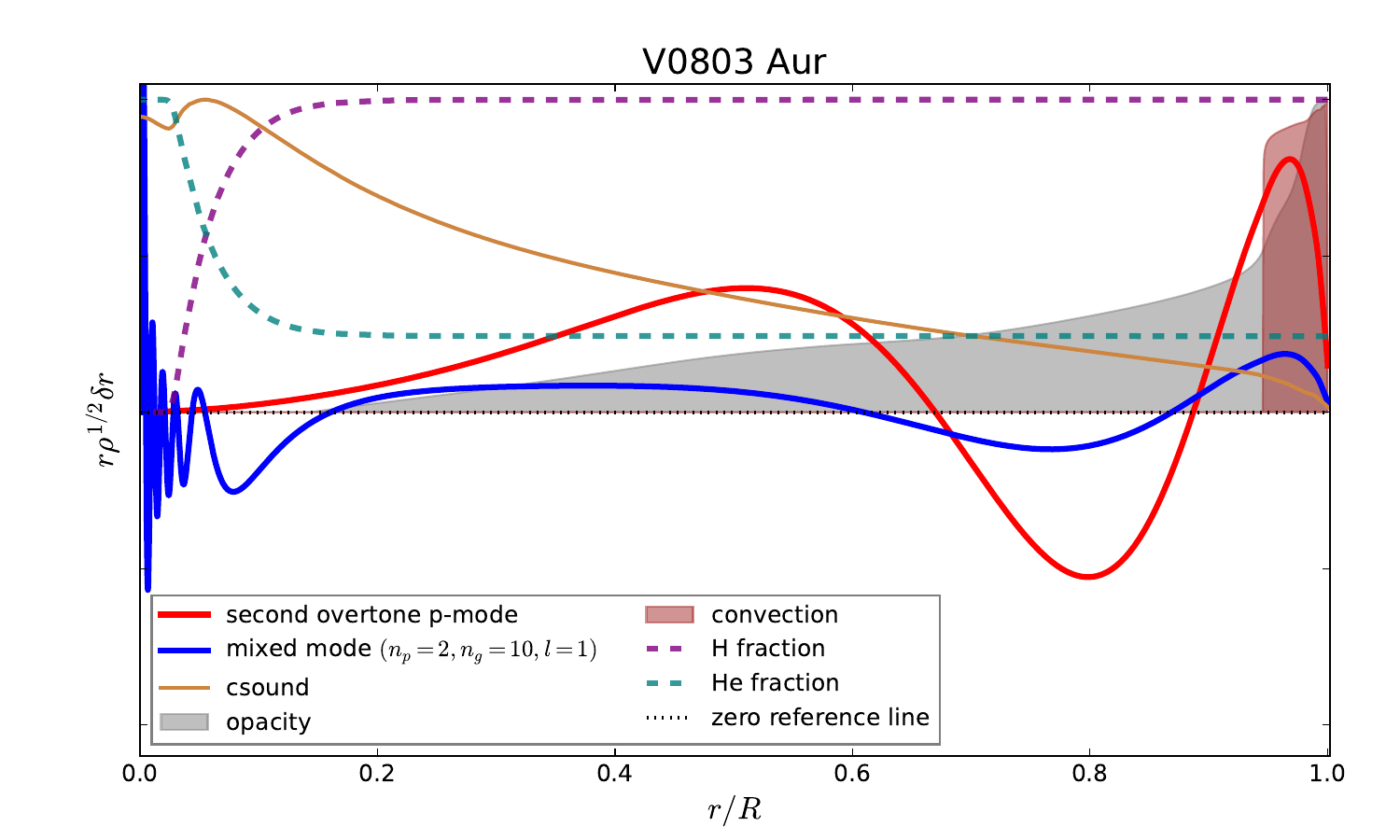}
  \includegraphics[width=0.49\textwidth]{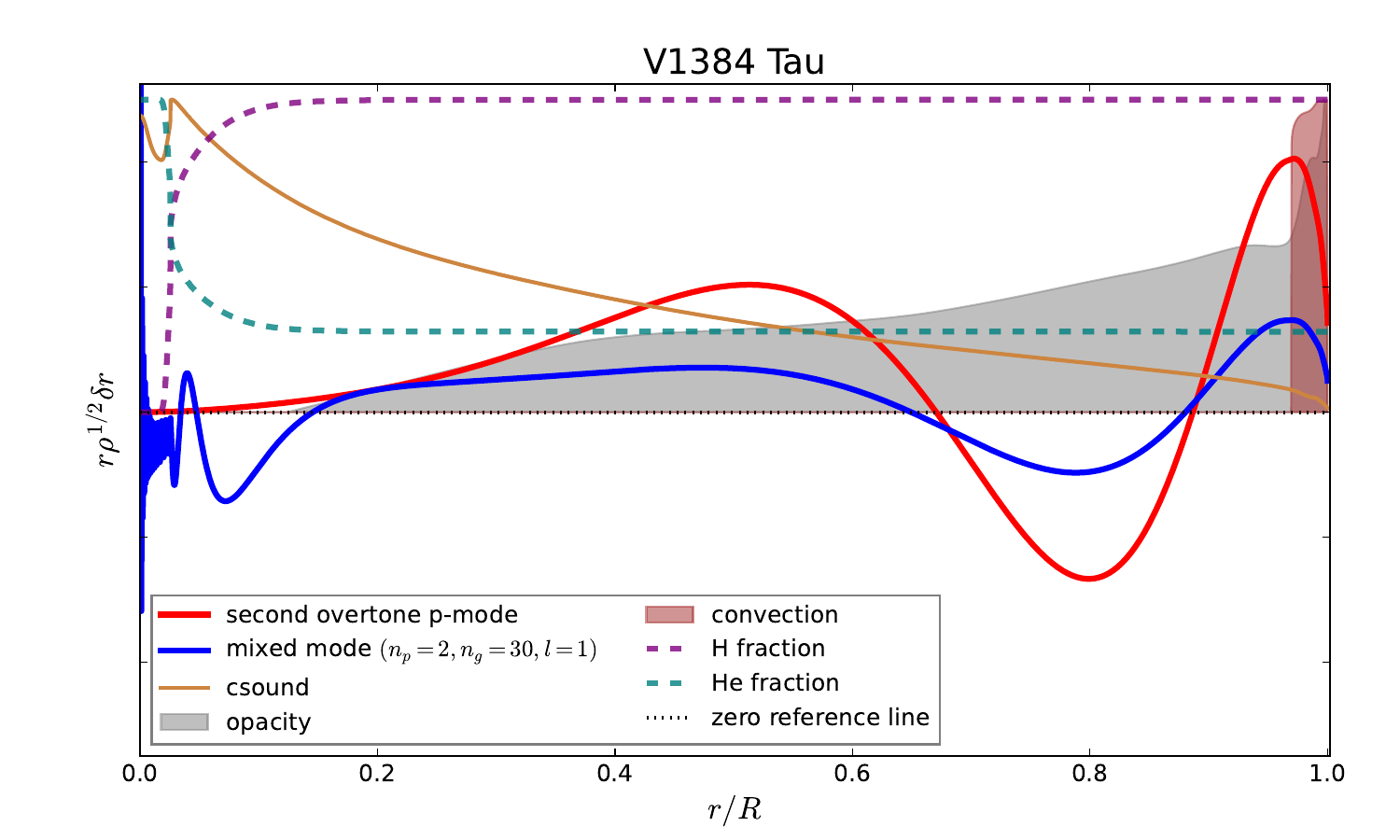}
  \includegraphics[width=0.49\textwidth]{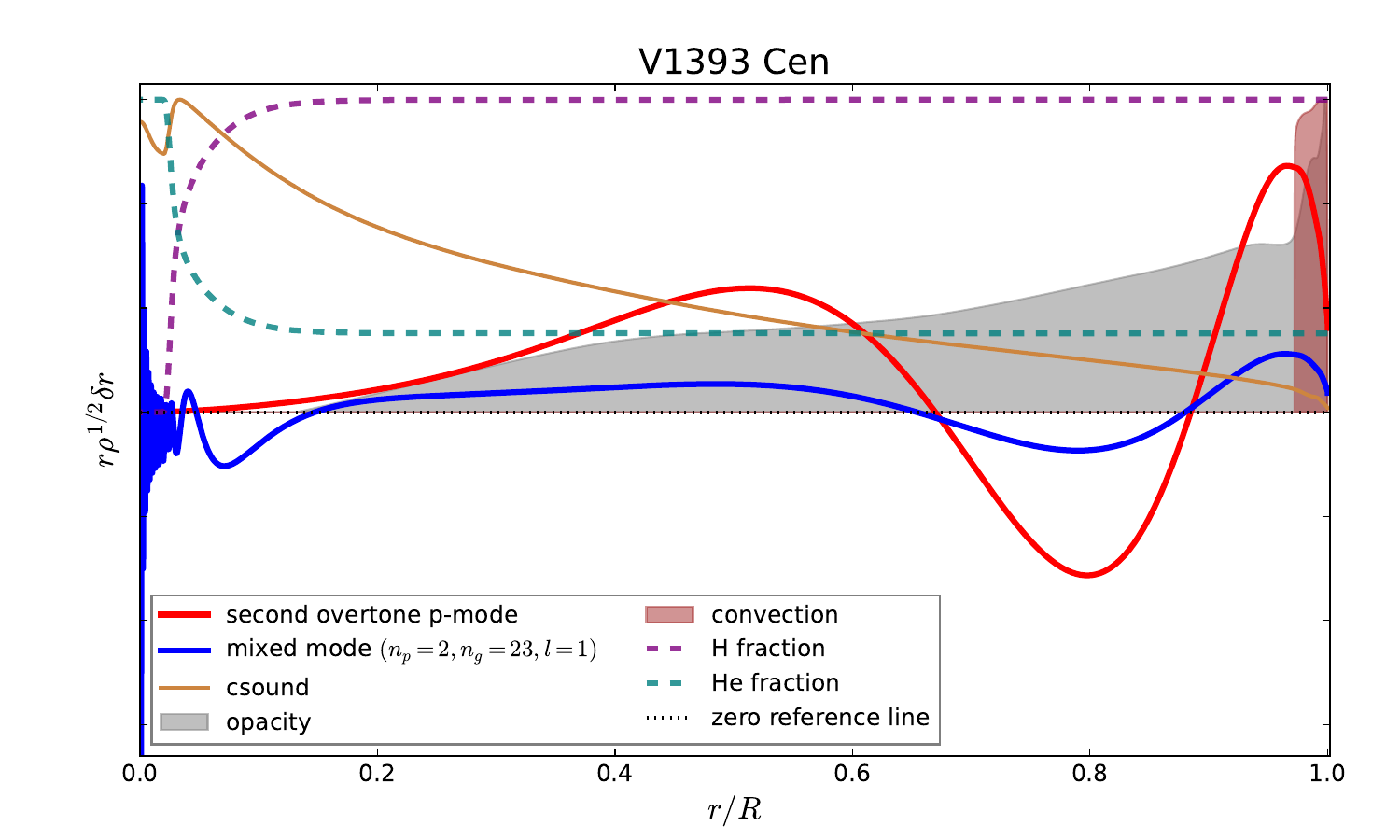}
  \caption{Radial displacement (represented by $r \rho^{1/2} \delta r$) of the related pulsation modes and some relevant profiles of the best-fit seismic models. Note that all the profiles are renormalized for a better presentation.}
  \label{fig:profile}
\end{figure*}

The H-R and Petersen diagrams with all evolutionary tracks and best-fit seismic models of the five HADS are shown in Figures \ref{fig:HR} and \ref{fig:petersen}, respectively.
The radial displacement (represented by $r \rho^{1/2} \delta r$) of the related pulsation modes (the second overtone and the p-g mixed mode) and some relevant profiles (including the sound speed (csound), opacity, convection velocity (convection), hydrogen and helium fractions (H fraction and He fraction)) of the best-fit seismic models are shown in Figure \ref{fig:profile}.

\end{appendix}
\end{document}